\documentclass[]{tWRM2e}

\usepackage{lineno}
\usepackage{color}
\definecolor{rouge}{rgb}{1,0,0}
\definecolor{bleu}{rgb}{0,0,1}
\definecolor{vert}{rgb}{0,0.5,0}

\begin{document}
\doi{10.1080/1745503YYxxxxxxxx}
 \issn{1745-5049}
\issnp{0278-1077}
\jvol{00} \jnum{00} \jyear{2012} \jmonth{January}

\markboth{M. Chekroun, L. Le Marrec, B. Lombard, J. Piraux}{Waves in Random and Complex Media}

\title{{\itshape Time-domain numerical simulations of multiple scattering\\ to extract elastic effective wavenumbers }}

\author{Mathieu Chekroun$^{\rm a}$
\vspace{6pt},
Lo\"{i}c Le Marrec$^{\rm b}$,
Bruno Lombard$^{\rm c}$$^{\ast}$\thanks{$^\ast$Corresponding author. Email: lombard@lma.cnrs-mrs.fr} and 
Jo\"{e}l Piraux$^{\rm c}$\\
\vspace{6pt}
$^{\rm a}${\em{Laboratoire d'Acoustique de l'Universit\'e du Maine, UMR CNRS 6613, avenue Olivier Messiaen, 72085 Le Mans, France}}; 
$^{\rm b}${\em{Institut de Recherche Math\'ematique de Rennes, UMR CNRS 6625, 263 avenue du G\'en\'eral Leclerc, 35042 Rennes, France}};
$^{\rm c}${\em{Laboratoire de M\'ecanique et d'Acoustique, UPR CNRS 7051, 31 chemin Joseph Aiguier, 13402 Marseille, France}}\\
\vspace{6pt}
\received{Received xx Xxx 2012}}

\maketitle

\begin{abstract}
Elastic wave propagation is studied in a heterogeneous 2-D medium consisting of an elastic matrix containing randomly distributed circular elastic inclusions. The aim of this study is to determine the effective wavenumbers when the incident wavelength is similar to the radius of the inclusions. A purely numerical methodology is presented, with which the limitations usually associated with low scatterer concentrations can be avoided. The elastodynamic equations are integrated by a fourth-order time-domain numerical scheme. An immersed interface method is used to accurately discretize the interfaces on a Cartesian grid. The effective field is extracted from the simulated data, and signal-processing tools are used to obtain the complex effective wavenumbers. The numerical reference solution thus-obtained can be used to check the validity of multiple scattering analytical models. The method is applied to the case of concrete. A parametric study is performed on longitudinal and transverse incident plane waves at various scatterers concentrations. The phase velocities and attenuations determined numerically are compared with predictions obtained with multiple scattering models, such as the Independent Scattering Approximation model, the Waterman-Truell model, and the more recent Conoir-Norris model.  
\bigskip

\bigskip
\end{abstract}  



\section{Introduction}\label{SecIntro}

We consider the propagation of elastic waves across a medium containing randomly distributed circular inclusions, the size of which is similar to that of the wavelength. The effective field, which is obtained by averaging the fields in all the possible disordered configurations, corresponds to that of waves propagating in an effective homogeneous medium. 

There exists three possible approaches for obtaining the effective wavenumbers (and equivalently, the effective phase velocity and attenuation):
\begin{itemize}
\item theoretical approach, based on multiple-scattering models such as the Foldy \cite{Foldy45}, Waterman-Truell \cite{WatermanTruell61}, and Fikioris-Waterman \cite{Fikioris64} models. It provides closed-form expressions useful in practical applications. The main assumption is that the scatterer concentration is low, i.e. typically less than 10 \% \cite{Martin06}. At higher concentrations, more sophisticated models developed in acoustics by Linton and Martin \cite{Linton05,Martin08}, and extended to elastodynamics by Conoir and Norris \cite{Linton05,Martin08,Conoir10,Luppe10}, are required. But, to our knowledge, no rigorous error estimate is available, and the limits of validity are not accurately known;
\item experimental approach \cite{Derode06}. It introduces no limitation about the concentration of scatterers, but it is very difficult to control accurately the various parameters involved: positions and geometries of scatterers, values of the physical parameters;
\item numerical approach \cite{Luneville10,Schubert04}. It allows a simpler control of the parameters and is fast; in practice, however, specific tools are required to perform efficient numerical computations and to render the numerical artifacts much smaller than the quantities of physical interest. 
\end{itemize}
The first aim of the present paper is to describe a numerical approach of this kind. The second aim is to highlight the efficiency of this methodology to explore the validity domain and limitations of analytical multiple scattering models \cite{Kawahara09}.  

For this purpose, we will proceed as follows. In section \ref{SecRandom}, the problem of obtaining random configurations is discussed; naive algorithms converge slowly when the scatterer concentrations are greater than 40 \%. The statistical behavior of the configurations is determined by performing a detailed analysis of the radial distribution function. In section \ref{SecSimu}, time-domain numerical methods are introduced. Elastodynamic equations are integrated using a high-order finite-difference time-domain scheme whose numerical artifacts are known in the case of a homogeneous medium. The discretization of the interfaces between the host matrix and the scatterers is a key issue: special care has to be taken here to prevent the interfaces to introduce large numerical artifacts for physical, geometrical and numerical reasons \cite{HDR-LOMBARD}. In section \ref{SecTDS}, signal-processing tools are applied to the simulated data, yielding the effective wavenumbers.

This numerical method is applied to a simple model of concrete, consisting of mortar containing composite inclusions \cite{Schubert04}. In section \ref{SecExp}, numerical experiments are performed at various inclusion concentrations (ranging from 3 \% to 60 \%), with longitudinal and transverse incident plane waves. Studies are performed to ensure that the averaged field obtained from a finite number of disordered configurations is representative of the theoretical effective field. 
In section \ref{SecRes}, wavenumbers are extracted from the simulated data. Comparisons are made with multiple-scattering models in terms of the concentration and the adimensional frequency. In particular, the advantages of recent developments over the traditional Waterman-Truell model are confirmed. In section \ref{SecConclu}, conclusions are drawn and some future lines of research are described. Technical details about the computation of theoretical wavenumbers are given in the appendix \ref{AppEff}.


\section{Random configurations}\label{SecRandom}

\subsection{Algorithm}\label{SecRandomAlgo}

\begin{figure}[htb]
\begin{center}
\begin{tabular}{c}
\includegraphics[scale=0.70]{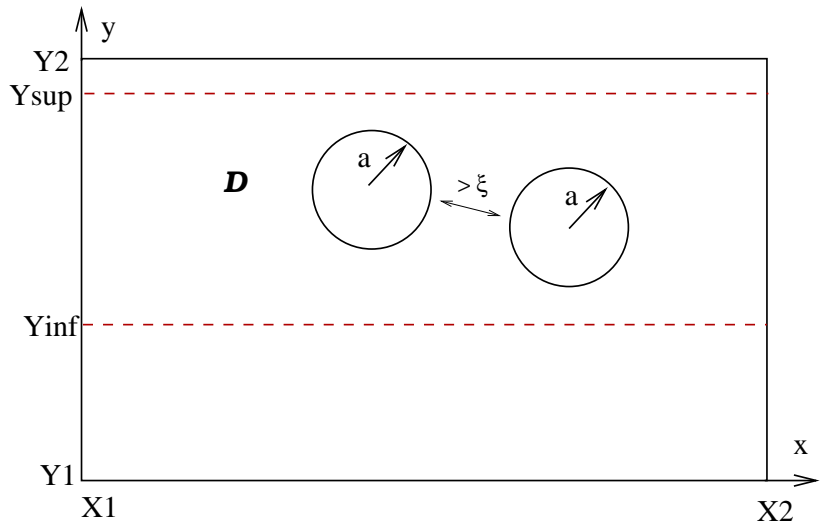}
\end{tabular}
\end{center}
\caption{Computational domain. Subdomain ${\cal D}$ containing the scatterers. A minimum exclusion distance $2\,a+\xi$ between the centers of the scatterers is assumed, where $\xi$ depends on the mesh size.}
\label{FigDomaine}
\end{figure}

Wave propagation is investigated in an infinite medium consisting of a matrix containing circular inclusions with a constant radius $a$, in the $x-y$ plane. Matrix and inclusions are in perfect bonded contact; they both consist of linear elastic isotropic homogeneous media. 

In practice, the time-domain numerical simulations are performed in a bounded computational domain $[X_1,\,X_2]\times[Y_1,\,Y_2]$. For this purpose, $N$ scatterers are introduced into the rectangular subdomain ${\cal D}=[X_1,\,X_2]\times[Y_{\inf},\,Y_{\sup}]$, where $Y_1<Y_{\inf}<Y_{\sup}<Y_2$ (figure~\ref{FigDomaine}). A minimum exclusion distance $2\,a+\xi$ between the centers of the scatterers is required by the numerical methods; $\xi$ increases with the mesh size (section \ref{SecSimuADER}). The $x$ and $y$ coordinates of the centers of circles $C_i$ ($i=1,\cdots,N$) are uniformly distributed in $[X_1,\,X_2]$ and $[Y_{\inf}+a,\,Y_{\sup}-a]$. Lastly, periodicity of the configuration is imposed along the $x$-axis, where the period is $X_2-X_1$. 

Various methods to simulate the $C_i$ may be found in the literature \cite{Siquiera95}. Here we propose two algorithms:

{\bf Algorithm 1}.

{\ttfamily
\noindent
$\triangleright$ choose $C_1$ randomly in ${\cal D}$;\\
$\triangleright$ for $i=2$ to $N$ do
\begin{enumerate}
\item[-] choose randomly $C_i$ in ${\cal D}$;
\item[-] if $C_iC_j\geq2\,a+\xi$ ($j=1,\cdots,i-1$) then $C_i$ is kept;
\item[-] otherwise choose another $C_i$.
\end{enumerate}
}
Algorithm 1 is very simple and gives quasi-uniform distributions. However, poor convergence is obtained at surface concentration $\phi$ greater than 30 \%, especially with large $\xi$. Surface concentrations $\phi$ greater than 50 \% are beyond the reach of this algorithm (\cite{TorquatoLivre}, p.67), even with $\xi=0$.

{\bf Algorithm 2}.

{\ttfamily
\noindent
$\triangleright$ a compact hexagonal packing pattern consisting of $N$ circles is initially introduced into ${\cal D}$. The side length of each hexagon is $2\,a+\xi$;\\
$\triangleright$ repeat until sufficiently uniform distributions are obtained:
\begin{enumerate}
\item[-] for $i=1$ to $N$ do
\begin{itemize}
\item perturb the position of $C_i$;
\item if $C_iC_j\geq2\,a+\xi$ ($j=1,\cdots,N$, $j\neq i$) then $C_i$ is kept.
\end{itemize}
\end{enumerate}
}
The mean values and variances of the $C_i$ coordinates are measured at each iteration. When their third decimal value no long varies, then the perturbation process is stopped. Configurations thus-obtained closely resemble to uniform distributions \cite{Abramovitz64}. Algorithm 2 can be used to reach surface concentrations up to roughly 66 \%.  

In practice, we recommend the use of algorithm 2 whatever the $\phi$. In order to obtain a sufficiently large number of disordered patterns, the algorithm selected is applied ${\cal N}$ times, which gives ${\cal N}$ independent configurations for each incident wave at each scatterer concentration (section \ref{SecRandomRDF}).


\subsection{Radial distribution function}\label{SecRandomRDF}

In this section, we examine numerically whether the above algorithms give uniform distributions. For this purpose, let us take the normalized radial distribution function (RDF)
$$
g(r)=p(r)\,\frac{\textstyle N}{\textstyle n_{0}},
$$
where $p$ is the conditional probability (appendix \ref{AppEff}), and $n_0$ is the number of scatterers per unit area. In an infinite statistically homogeneous domain, $g\to 1$ as $r\to \infty$. The RDF is calculated numerically by counting the number $n(r)$ of inclusion centers present in a circular ring with radius $r$ and thickness $\Delta r$: 
\begin{equation}
g(r)=\frac{n(r)}{n_{0}}\frac{\textstyle 1}{\textstyle 2\,\pi\, r\, \Delta r},
\hspace{1cm} 2\,a+\xi\leq r\leq r_{max},
\label{RDF}
\end{equation}
where $r_{max}$, the distance to the nearest boundary of $\mathcal{D}$, is used in order to prevent bounding effects. In practice, we take $\Delta r=a\,/\,20$. The RDF is calculated for each inclusion in the simulation domain in order to obtain a representative value. In the case of dilute media, the number $N$ of inclusions is too low to obtain a smooth curve, and it is not possible to determine the typical behavior of the RDF. The number of configurations ${\cal N}$ is increased until a standard deviation on $g(r)$ of around 5\% is obtained when $7\leq r/a\leq 10$: in this range, the RDF is stabilized at $1$. The parameters of these calculations are given in table \ref{tab:RDFlisse}.

\begin{table}[htb]
\begin{center}
\begin{tabular}{c| c c c c c c c c c c  }
$\phi$ &$6\%$ &$12\%$ &$18\%$ &$24\%$ &$30\%$ &$36\%$ &$42\%$ &$48\%$ &$54\%$ &$60\%$ \\
\hline
${\cal N}$
&$100$ &$30$ &$30$ &$20$ &$10$ &$10$ &$10$ &$3$ &$3$ &$3$ \\
$N$
& 573  & 1145 & 1719 & 2292 & 2865 & 3438 & 4011 & 4584 & 5157 & 5730
\end{tabular}
\caption{Parameters for the RDF calculations (\ref{RDF}): surface concentration $\phi$, number of configurations ${\cal N}$, number of inclusions $N$.}
\label{tab:RDFlisse}
\end{center}
\end{table}

\begin{figure}[htb]
\begin{center}
\begin{tabular}{cc}
\includegraphics[scale=0.42]{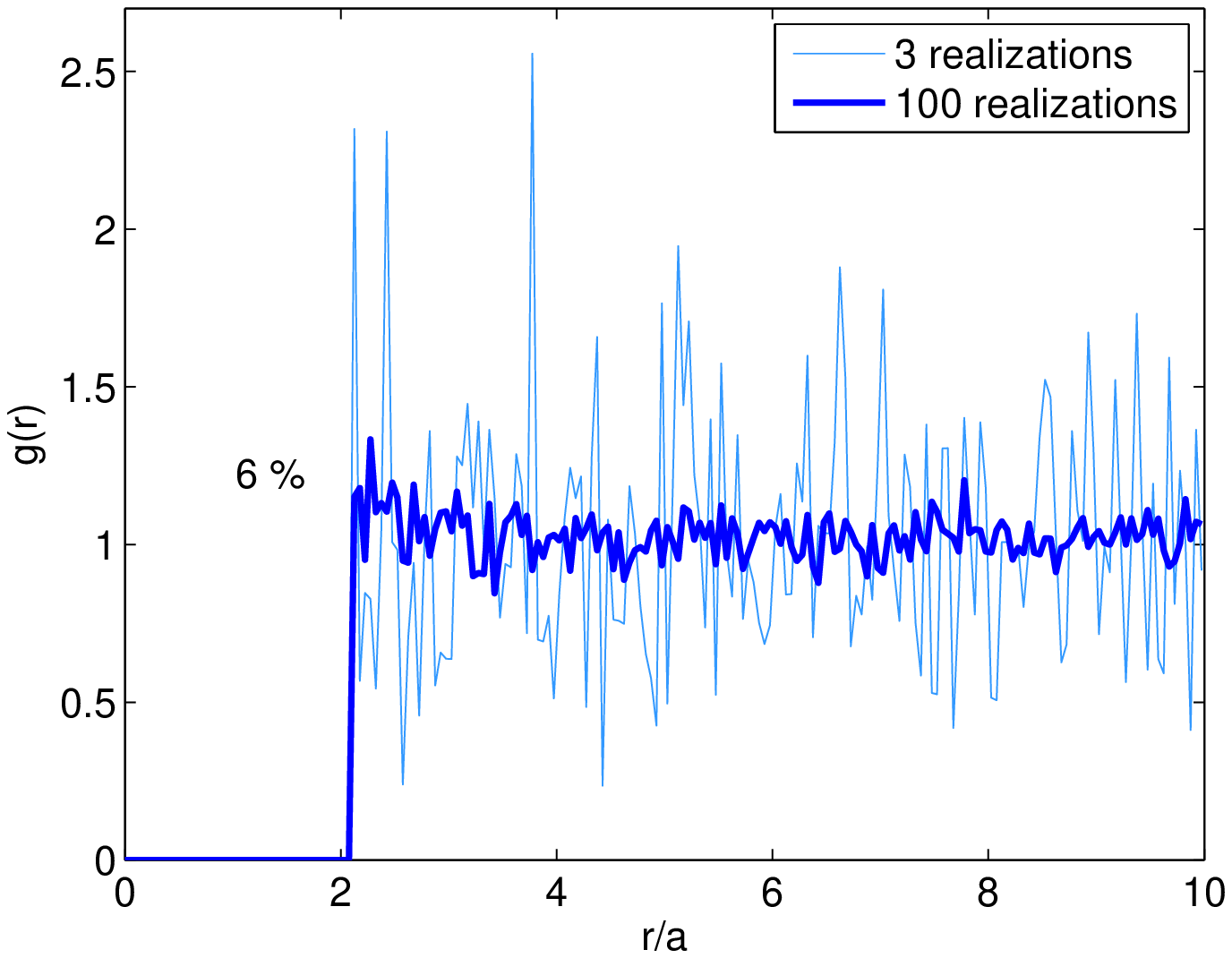} &
\includegraphics[scale=0.42]{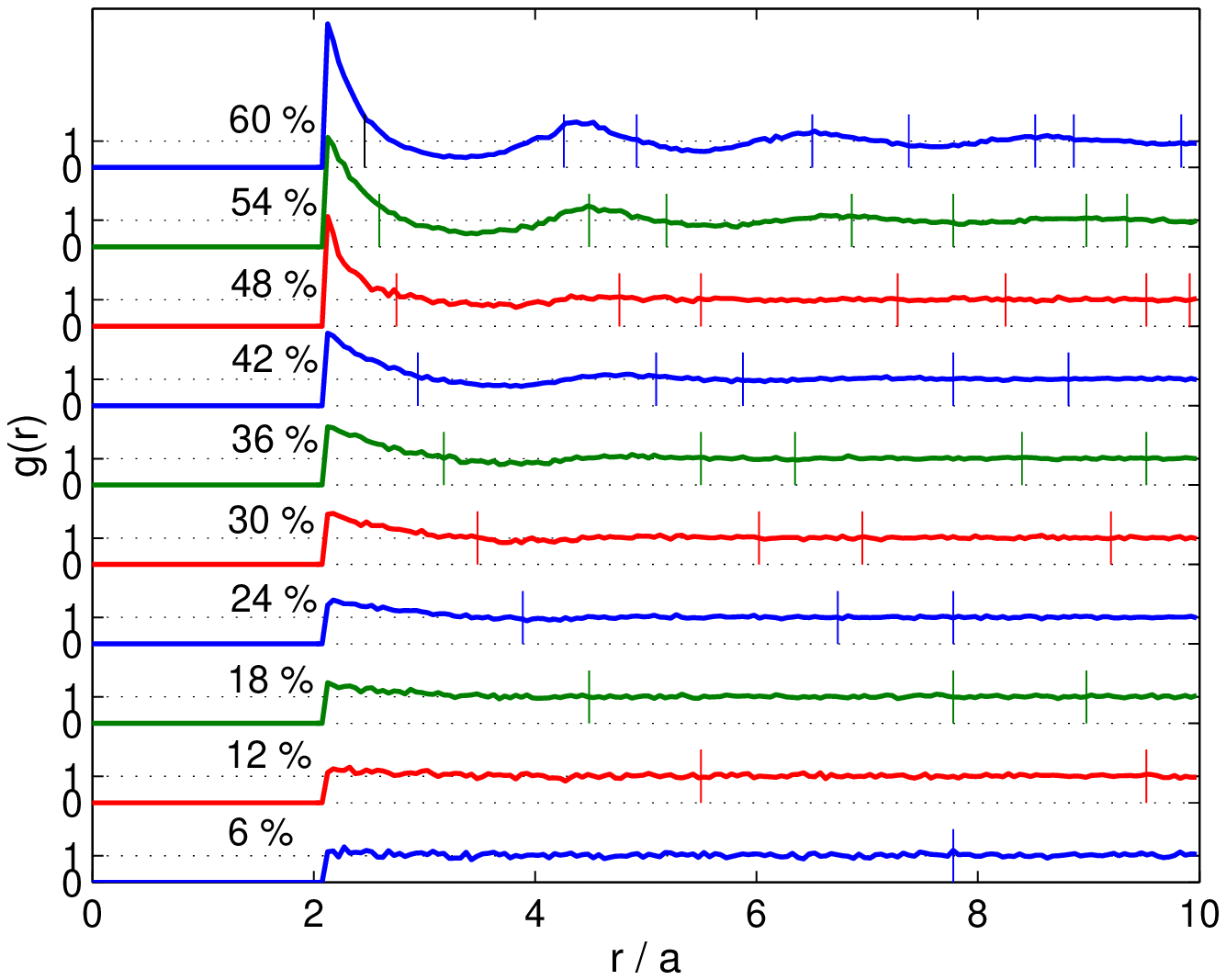}
\end{tabular}
\end{center}
\caption{Radial distribution function $g(r)$ at various concentrations $\phi$, calculated with $\Delta r=a/20$ in (\ref{RDF}). Left: RDF at $\phi=6\%$, with ${\cal N}=3$ and ${\cal N}=100$. Right: RDF at various surface concentrations $\phi$, calculated with the parameters given in table \ref{tab:RDFlisse}; the thin vertical lines indicate the positions of the Dirac distributions in the case of a hexagonal lattice.}
\label{fig:distrib}
\end{figure}

As can be seen in figure \ref{fig:distrib}, the RDF depends greatly on the concentration of the inclusions. At low concentration ($\phi\lesssim 10\,\%$), the RDF can be satisfactorily approximated by a Heaviside function if $N$ is sufficiently large. In other words, the conditional probability is uniform in this case. At higher concentrations, the local density of the neighbors in the vicinity of a given inclusion is increased. If $\phi\lesssim 30\,\%$, this local increase is proportional to $\phi$, as predicted by the virial expansion \cite{Varadan89}, whereas the conditional probability is uniform if $r \gtrsim 4\,a$. With more densely packed media ($\phi\gtrsim 40\,\%$), attenuated oscillations occur periodically. As the concentration increases, the oscillations occur farther away from the inclusion, and the period decreases. Our computation of theoretical effective wavenumbers will take into account the distance of exclusion between scatterers (appendix \ref{AppEff}). But incorporation of non-uniform RDF is beyond the scope of this paper; references on that topic may be found in \cite{Siquiera95,Caleap12}.

The random arrangement of circular inclusions has been studied in detail in many contexts. The behavior obtained here is in agreement with the results of statistical analyses. For further information about these distributions, see \cite{Torquato02}. The most compact 2D-arrangement corresponds to a hexagonal lattice. The RDF of this crystal is composed of Dirac distributions. As shown in figure \ref{fig:distrib} (right), the positions of the peaks in the simulated RDF do not correspond to the position of the Dirac distributions in the lattice. Even in the case of densely packed media, the heterogeneous structure cannot be approximated by a pseudo-periodic medium.


\section{Time-domain simulations}\label{SecSimu}

\subsection{Integration of elastodynamic equations}\label{SecSimuADER}

A velocity-stress formulation of elastodynamics is followed. The physical parameters are the density $\rho$, the speeds of longitudinal waves $c_L$ and of transverse waves $c_T$. The unknown are the horizontal and vertical velocity ($v_x$, $v_y$) and the independent components of the stress tensor ($\sigma_{xx}$, $\sigma_{xy}$, $\sigma_{yy}$). One has to solve the first-order linear hyperbolic system
\begin{equation}
\frac{\textstyle \partial}{\textstyle \partial\,t}\,{\bf U}+{\bf A}\,\frac{\textstyle \partial}{\textstyle \partial\,x}\,{\bf U}+{\bf B}\,\frac{\textstyle \partial}{\textstyle \partial\,y}\,{\bf U}={\bf 0},
\label{LC}
\end{equation}
where ${\bf U}=(v_x,\,v_y,\,\sigma_{xx},\,\sigma_{xy},\,\sigma_{yy})^T$, and ${\bf A}$ and ${\bf B}$ are $5\times 5$ matrices depending on the physical parameters. The system (\ref{LC}) is solved on a uniform Cartesian grid of $N_x\times N_y$ nodes, with mesh sizes $\Delta x=(X_2-X_1)\,/\,N_x$ and $\Delta y=(Y_2-Y_1)\,/\,N_y$, and a time step $\Delta t$. In practice, $\Delta x=\Delta y$. An explicit two-step finite-difference ADER (Arbitrary DERivatives) scheme is used, giving fourth-order accuracy in both space and time \cite{Dumbser04,Kaser06}. With this scheme, the minimal extra distance between two scatterers is $\xi=3\,\Delta x$ (section \ref{SecRandomAlgo}). The CFL limit of stability is 
\begin{equation}
\theta=c_{\max}\,\frac{\textstyle \Delta t}{\textstyle \Delta x}\leq 1,
\label{CFL}
\end{equation}
where $c_{\max}$ is the maximum speed of the waves in the domain. A plane wave analysis of this scheme has been performed in the case of a homogeneous medium \cite{HDR-LOMBARD}, in terms of $\theta$ and $G=\Delta x/\lambda$, $G\in]0,\,0.5]$, where $\lambda$ is the wavelength. The maximum artifacts are obtained when the direction of the propagation coincides with the grid axes, that is in the case of 1-D configurations. In this case, one has 
\begin{equation}
\begin{array}{l}
\displaystyle
q(\theta,\,G)=1-\frac{\textstyle 2\,\pi^4}{\textstyle 15}\,(\theta^2-1)\,(\theta^2-4)\,G^4+\mathcal{O}(G^6),\\
[8pt]
\displaystyle
\alpha(\theta,\,G)=\frac{\textstyle 4\,\pi^6}{\textstyle 9}\,\theta\,(\theta^2-1)\,(\theta^2-4)\,G^6+\mathcal{O}(G^8),
\label{AnaNum}
\end{array}
\end{equation}
where $q$ is the ratio between the exact and discrete phase velocities, and $\alpha$ is the discrete attenuation \cite{Strickwerda89}. The relations (\ref{AnaNum}) have crucial effects on the accuracy of the simulations, because they bound the numerical artifacts in homogeneous media. 


\subsection{Discretization of interfaces}\label{SecSimuESIM}

Three classes of drawbacks are classically associated with interfaces in finite-difference schemes. First, since the geometrical description of arbitrarily-shaped interfaces is poor, spurious diffractions are generated. Secondly, since the jump conditions are not enforced numerically, convergence may occur towards a non-physical solution. Lastly, non-smoothness of the solution across interfaces decreases the accuracy of the scheme, leading to spurious oscillations and even to instabilities. These three drawbacks increase with the scatterer concentration and preclude the use of  simulations as metrological tools in highly heterogeneous media. An alternative strategy consists in using numerical methods with unstructured meshes, such as finite-element methods, Galerkin discontinuous methods, and spectral element methods \cite{Komatitsch99,Luneville10}. But the computational cost of these methods would be much higher due to the meshing, and the stability condition (\ref{CFL}) is penalized.

To overcome these drawbacks, we use a $r$-th order immersed interface method \cite{Lombard04,HDR-LOMBARD}. This numerical method modifies the ADER scheme at grid points close to the interfaces, based on the jump conditions up to the $r$-th order, the elastodynamic equations, and the Beltrami equations. This procedure associates the efficiency of Cartesian grid methods and the accuracy of an interface meshing. The work is mainly carried out during a preprocessing step, before the numerical integration. At each time step, ${\cal O}({\cal L}\,/\,\Delta x)$ matrix-vector products are done, where $\mathcal{L}$ is the total perimeter of the interfaces, and the matrices are small, typically $5 \times 100$. The results are then injected into the scheme. After optimizing the codes, the additional CPU time required by the immersed interface method can be made negligible in comparison with the CPU time required by the scheme (less than 1\%).


\subsection{Intensive computing}\label{SecSimuParal}

To obtain reliable effective wavenumbers values, the numerical methods used must meet the following specifications:
\begin{itemize}
\item large computational domains, such as grids consisting of $10^4\times 10^4$ nodes, involving 10\,GB of data;
\item long integration times, consisting for example of $10^4$ time steps;
\item several simulations, so as to increase the number of independent disordered configurations, to ${\cal N}=3$, for example;
\item processing a large number of scatterers with the immersed interface method: 1500 interfaces when $\phi=48$ \%, for example;
\item performing many simulations in the parametric studies, e.g. in terms of the scatterer concentration or the incident wave polarizations.
\end{itemize}

To meet these specifications, the computer codes are parallelized. Domain decomposition is performed in the $x$-direction, associating each slide with one computational process. All the slices have the same size and contain approximately the same number of inclusions, so that the computational cost of each process is roughly the same. After each time step, data are exchanged between neighboring processes.

In practice, the simulations presented in sections \ref{SecExp} and \ref{SecRes} were performed on a cluster of 4 PC bi-processor quadricores, amounting to 32 processes. The optimum speed-up 32 was reached. The communication time between processes was negligible in comparison with the computational cost of each process. After parallelizing, the configurations investigated in section \ref{SecExp} required 1 hour of preprocessing (due to the use of the immersed interface method) and 24 hours of integration (due to the use of the ADER scheme).


\section{Data processing}\label{SecTDS}

\subsection{Numerical coherent field}\label{SecTDS1}

At each time step, the components of ${\bf U}$ have to be stored inside the subdomain containing the inclusions. For this purpose, a uniform network consisting of $N_l$ lines and $N_c$ columns of receivers is placed in the subdomain ${\cal D}$. The position of the receivers is given by $(x_i=X_1+i\,\Delta_c,\,y_j=Y_{\inf}+j\,\Delta_l)$, where $i=0,\hdots,\,N_c-1$ and $j=0,\hdots,\,N_l-1$. The bottom and top receivers are sufficiently far from the sides of the computational domain to prevent spurious effects from being recorded. These columns of receivers are visible in figure \ref{FigCarteInit}-(a).

The acquisition setup has to meet some specifications in order to prevent the occurrence of aliasing and low resolution problems \cite{Forbriger03}. Aliasing occurs when the distance $\Delta_l$ is larger than the shortest wavelength under consideration, while the resolution is limited by the total length $N_l\,\Delta_l$ of the acquisition setup (i.e. the distance between the first and last receivers). 

The field recorded on each array (each column of receivers) corresponds to a field propagating in a given disordered configuration. Summing the time histories of these $N_c$ arrays in all the ${\cal N}$ simulations gives a coherent field propagating in the $y$ direction. The relationship between the coherent field and the effective field will be discussed in section \ref{SecExpConverge}.

The polarization of the coherent field is the same as the polarization of the incident plane wave. In the following section, we will therefore deal with either a $L$ (longitudinal) or a $T$ (transverse) incident wave.


\subsection{Extraction of the coherent wavenumbers}\label{SecTDS2}

The coherent phase velocity $c(\omega)$ is computed by applying a $\mathfrak{p}\!-\!\omega$ transform to the space-time data on the coherent field, where $\mathfrak{p}$ is the slowness of the waves ($\mathfrak{p}=1\,/\,c$) and $\omega$ is the angular frequency \cite{McMechan81,Mokhtar88}. The time Fourier transform of the coherent field $s(y_j,\,\omega)$ is denoted by
\begin{equation}
s(y_j,\,\omega)\,=\,A(y_j,\,\omega)\,\mathrm{e}^{-i\,\omega\,\mathfrak{p}_0(\omega)\,y_j},
\label{PW1}
\end{equation}
where $A(y_j,\,\omega)$ is the amplitude spectrum at $y_j$, and $\mathfrak{p}_0(\omega)$ needs to be determined. A $\mathfrak{p}\!-\!\omega$ stack quantity $\hat{s}(\mathfrak{p},\,\omega)$ is then defined as
\begin{equation}
\hat{s}(\mathfrak{p},\,\omega)=\sum_{j=0}^{N_l-1} A(y_j,\,\omega)\,e^{i\,\omega (\mathfrak{p}-\mathfrak{p}_0(\omega))\,y_j}.
\label{PW2}
\end{equation}
$\hat{s}(\mathfrak{p},\,\omega)$ is computed at several $\mathfrak{p}$ values. Given $\omega$, the maximum value of the modulus $|\hat{s}(\mathfrak{p},\,\omega)|$ is reached at $\mathfrak{p}=\mathfrak{p}_0(\omega)=1\,/\,c(\omega)$. The phase velocity dispersion curve is then obtained by taking the maximum locus on the 2-D map $|\hat{s}(\mathfrak{p},\,\omega)|$. An error estimate is also deduced \cite{Herrmann02}.

Let us now examine the attenuation of the coherent field. This quantity is estimated from the decrease in the amplitude spectrum of the coherent field during the propagation of the waves. In the frequency domain, the amplitude $A(y_j,\,\omega)$ in (\ref{PW1})-(\ref{PW2}) is assumed to satisfy an exponential decay with distance 
\begin{equation}
A(y_j,\,\omega)=A_0(\omega)\,\mathrm{e}^{-\alpha(\omega)\,(y_j - y_0)},
\label{Alpha}
\end{equation}
where $A_0(\omega)$ is the amplitude of $s$ at the first receiver located at the offset $y_0$, and $\alpha(\omega)$ is the attenuation factor. Although two points suffice to be able to calculate $\alpha$, the slope of a least-squares fit of $\ln (A(y_j,\,\omega))$ over the whole range of reception gives more accurate results. An error estimate is also deduced. 


\section{Numerical experiments}\label{SecExp}

\subsection{Validation}\label{SecExpValid}

\begin{figure}[htb]
\begin{center}
\begin{tabular}{cc}
(a) & (b)\\
\includegraphics[width=6.7cm]{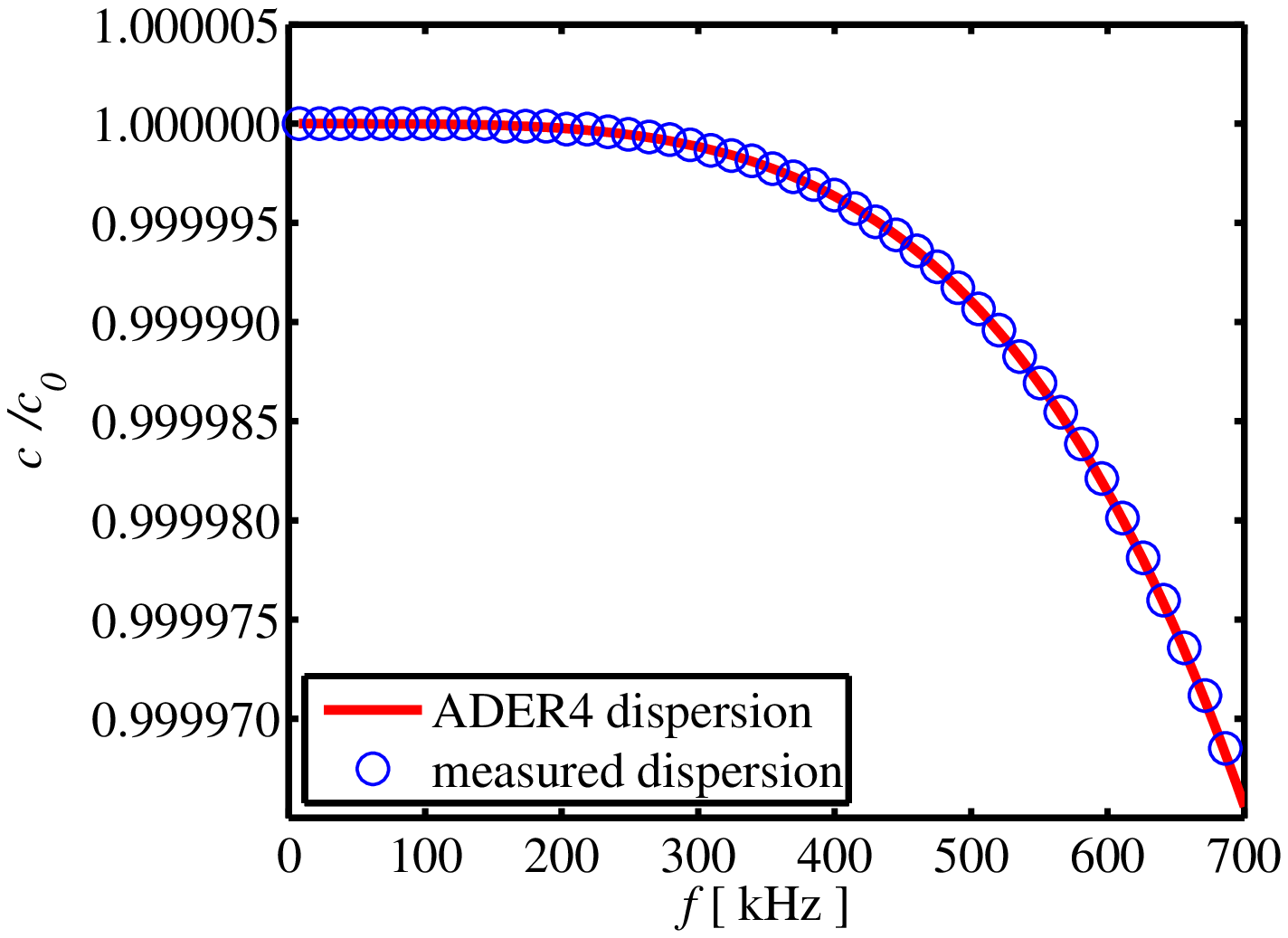} &
\includegraphics[width=6.7cm]{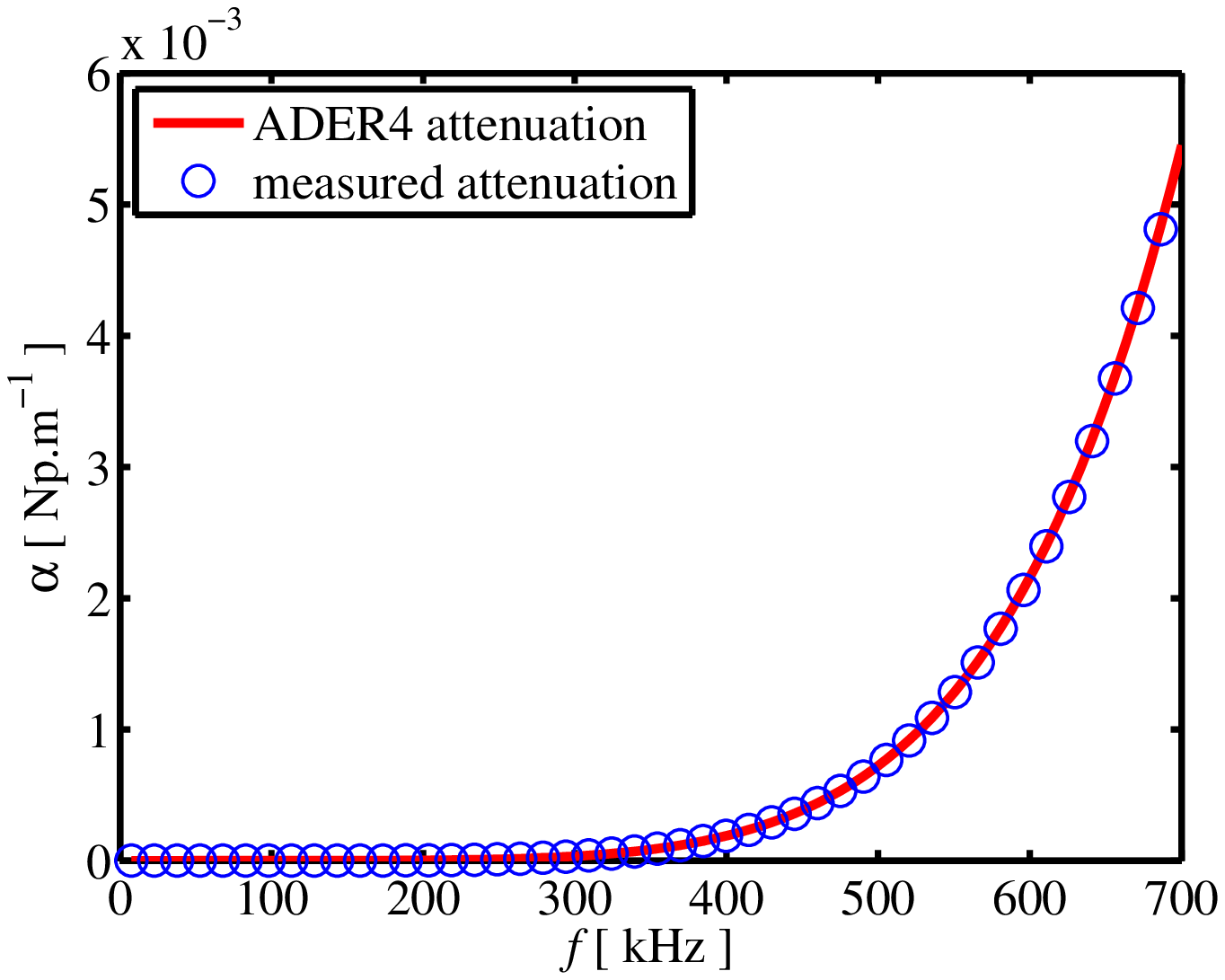}
\end{tabular}
\end{center}
\caption{Validation test 1/2. Numerical dispersion (a) and numerical attenuation (b) in a 1D homogeneous medium: analytical values (\ref{AnaNum}) in red lines, and measured values in blue circles.}
\label{FigValid1D}
\end{figure}

In the first test, transverse wave propagation was simulated in a homogeneous 1-D cement matrix. The dispersion and attenuation measured were  due only to numerical artifacts occurring in the ADER scheme. Comparisons between the theoretical (\ref{AnaNum}) and measured dispersion and attenuation values is made in figure \ref{FigValid1D}. The error between the theoretical and measured curves is less than $10^{-3}$\% in the frequency range of interest. The signal processing tools used and the acquisition setup chosen are therefore suitable for accurately assessing the dispersion and the attenuation, and the risk of adding significant signal processing artifacts is thus avoided. 

\begin{figure}[htb]
\begin{center}
\begin{tabular}{cc}
(a) & (b)\\
\includegraphics[scale=0.33]{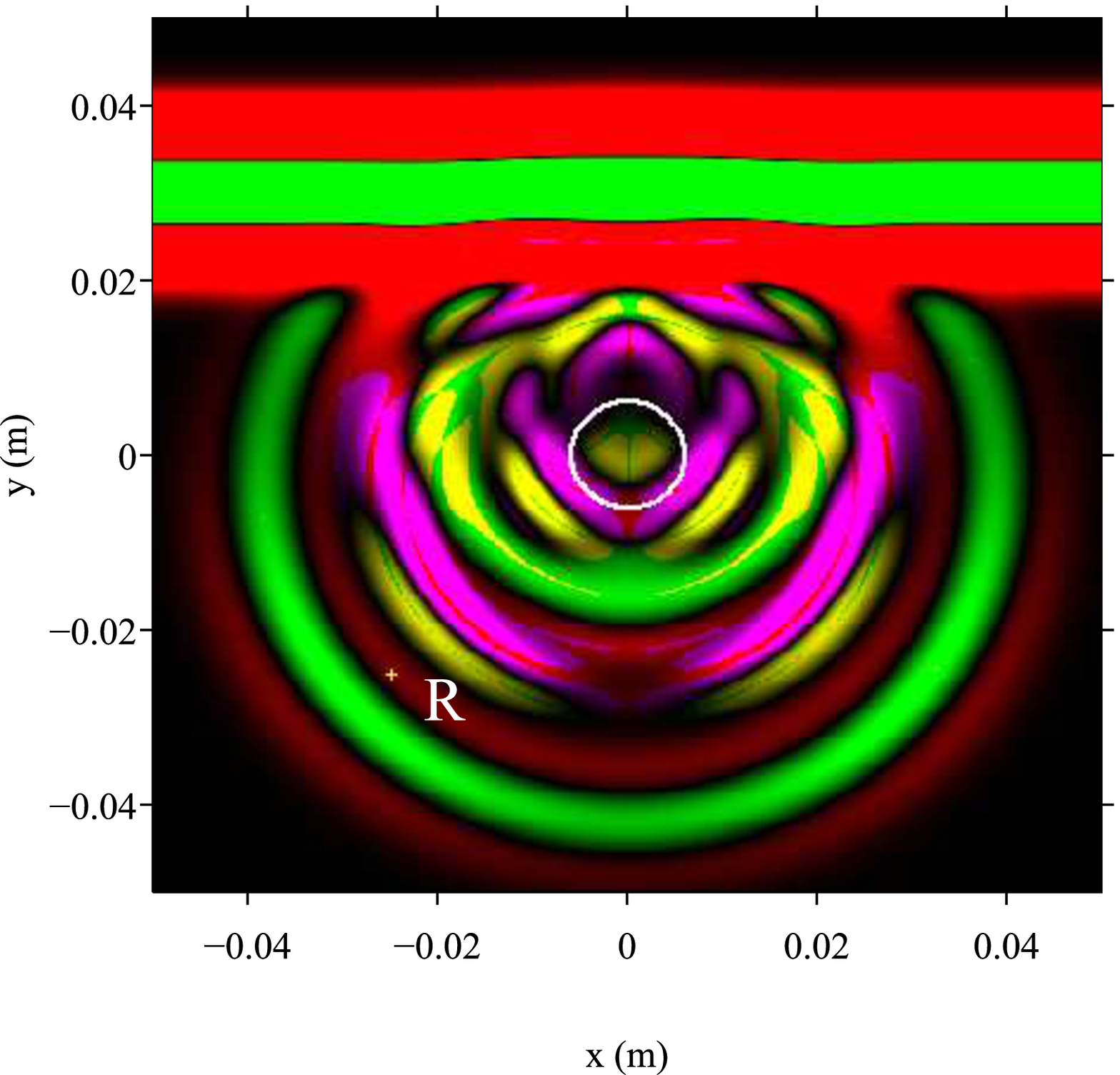} &
\includegraphics[scale=0.33]{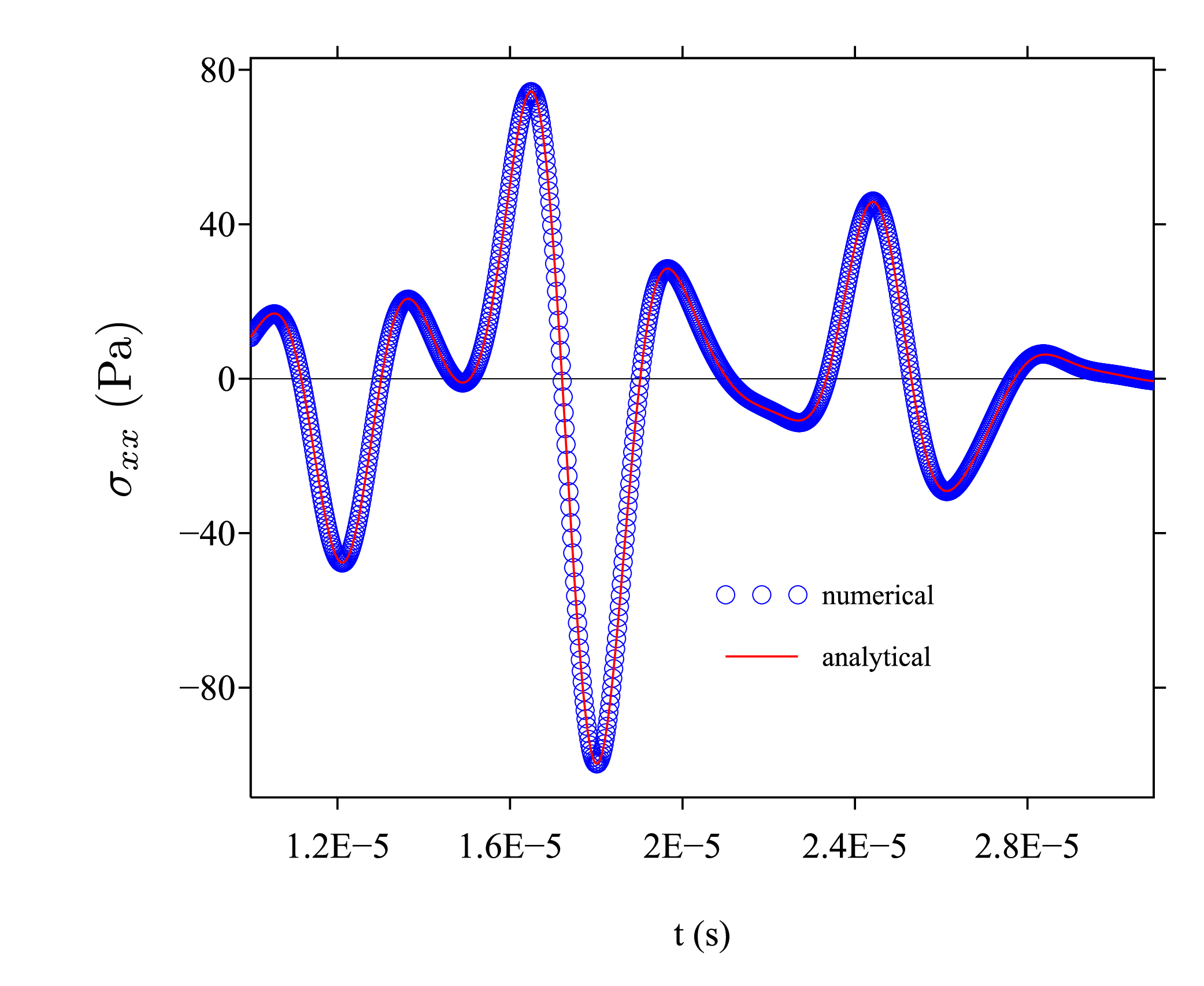}
\end{tabular}
\end{center}
\caption{Validation test 2/2. Snapshot and time-history of $\sigma_{xx}$: comparison between the numerical and the analytical solutions at the receiver R. In (a), the green-red palette and yellow-magenta palette denote $L$ waves and $T$ waves, respectively.}
\label{FigValid2D}
\end{figure}

In the second test, the wave propagation was simulated in a 2-D medium with a single scatterer centered at $(0,\,0)$. The source was the plane compressional wave described in section \ref{SecExp}. The diffracted fields were stored in the receiver R at $(-0.25,\,-0.25)$. Figure \ref{FigValid2D} gives a snapshot of the stress $\sigma_{xx}$ after 600 time steps (a) and compares the numerical and analytical values of $\sigma_{xx}$ during 1400 time steps (b). The exact solution was computed by performing standard Fourier-Bessel decompositions. The excellent agreement observed confirms the validity of both the ADER scheme and the immersed interface method.


\subsection{Numerical setup}\label{SecExpSetup}

\begin{figure}[htbp]
\begin{center}
\begin{tabular}{c}
(a)\\
\includegraphics[scale=0.4]{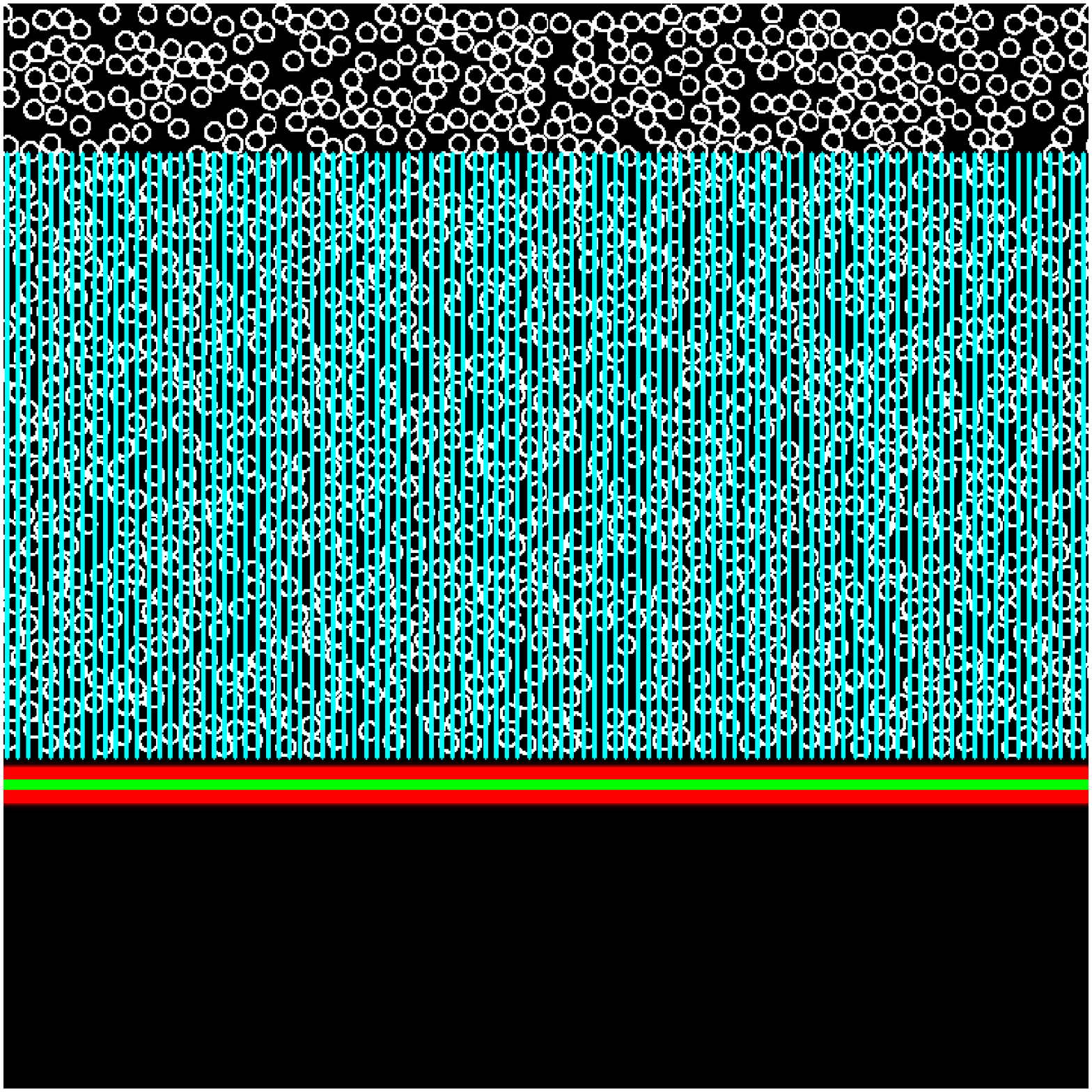}\\
\\
(b)\\
\includegraphics[scale=0.4]{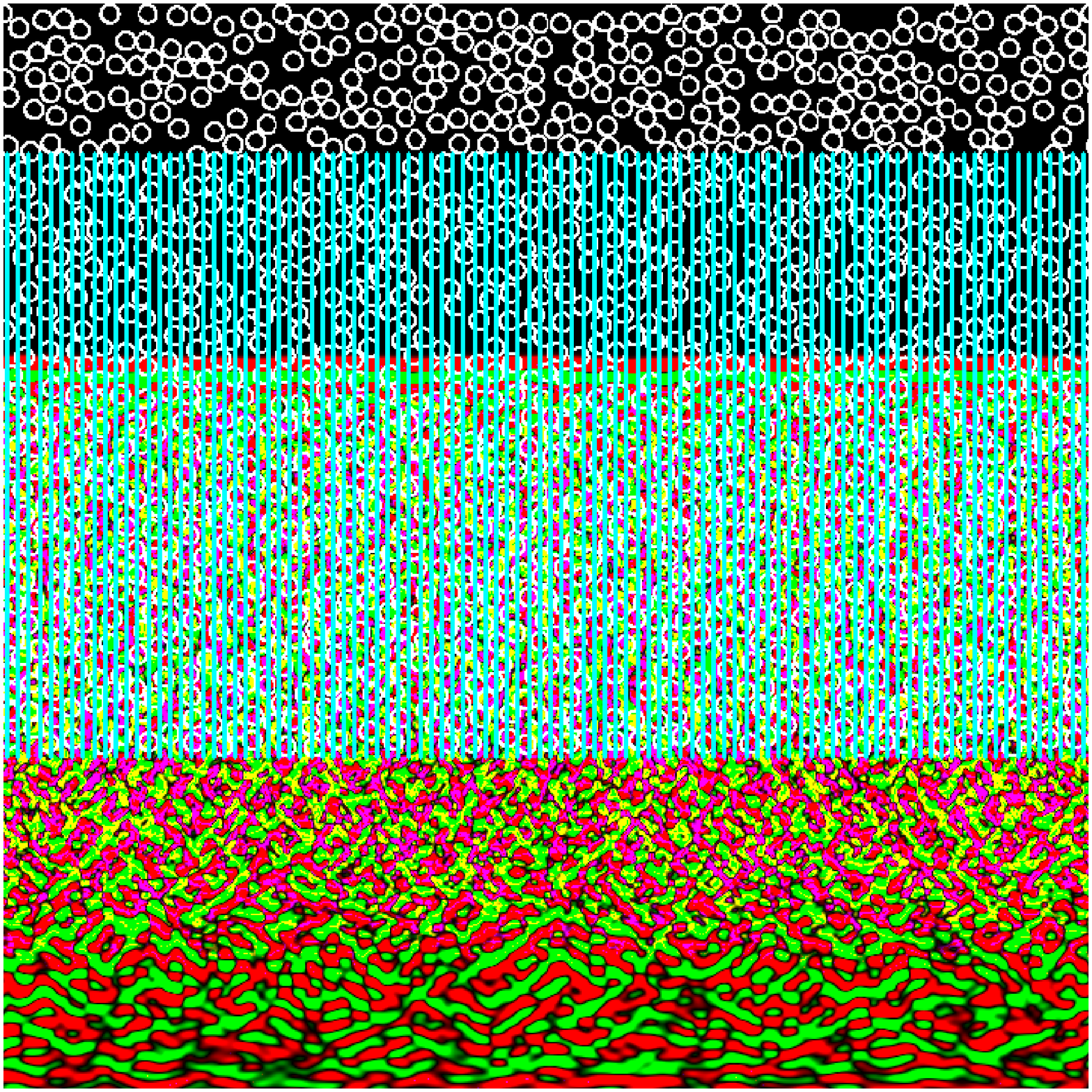}
\end{tabular}
\end{center}
\caption{Incident L-wave, concentration $\phi=42$ \%, initial instant (a) and after 3000 time steps (b). The vertical columns denote the positions of the receivers.}
\label{FigCarteInit}
\end{figure}

The numerical method presented in sections \ref{SecRandom} to \ref{SecTDS} is now applied to some physically relevant configurations. A simple model of concrete is studied, where circular aggregates with a radius $a=6$\,mm are embedded in a homogeneous cement matrix. The physical parameters are
\begin{equation}\nonumber
\begin{array}{rcl}
\displaystyle (\rho_{0},\,c_{0,L},\,c_{0,T})&=&\displaystyle
(2050\,\mbox{kg.m}^{-3},\,3950\,\mbox{m.s}^{-1},\,2250\,\mbox{m.s}^{-1}) \mbox{ in the cement  matrix},\\
[8pt]
\displaystyle (\rho_{1},\,c_{1,L},\,c_{1,T})&=&\displaystyle
(2610\,\mbox{kg.m}^{-3},\,4300\,\mbox{m.s}^{-1},\,2470\,\mbox{m.s}^{-1}) \mbox{ in the aggregates}.
\end{array}
\end{equation} 
A parametric study is performed in terms of the concentration, from $\phi=3$\% to 60\%. The domain of investigation presented in table \ref{TabDomaine} is discretized on $N_x \times N_y=7200\times 7200$ nodes, hence $\Delta x=\Delta y=10^{-4}$ m. The CFL number (\ref{CFL}) is $\theta=0.95$, giving $\Delta t=2.21\,10^{-8}$ s. A third-order immersed interface method is implemented ($r=3$ in section \ref{SecSimuESIM}). The source is a plane longitudinal (L) or transverse (T) wave propagating along the $y$-axis, initially outside the domain ${\cal D}$ (figures \ref{FigDomaine} and \ref{FigCarteInit}-a). The time evolution of the source is a Ricker with a central frequency 250\,kHz. The frequencies of interest range between 50 kHz and 600 kHz. 

\begin{table}[htb]
\begin{center}
\begin{tabular}{cccccc}
$X_1$ & $X_2$ & $Y_1$ & $Y_2$ & $Y_{\inf}$ & $Y_{\sup}$ \\
\hline
-0.36 & +0.36 & -0.02 & +0.7 & +0.2      & +0.6
\end{tabular}
\caption{Coordinates of the physical domain and those of the subdomain ${\cal D}$ (section \ref{SecRandom}), in meters.}
\label{TabDomaine}
\end{center}
\end{table} 

The acquisition network contains $N_c=100$ columns and $N_l=400$ lines, with the spacing $\Delta_l=0.001$ m and $\Delta_c=0.0072$ m, respectively. Each column is an array of receivers that follows the wave propagation in a particular realization of disorder. Performing ${\cal N}=3$ simulations yields $3\times 100$ independent disordered configurations. This acquisition setup gives the following bounds on the standard errors: from 2\,m.s$^{-1}$ at $f=50$ kHz to 0.2\,m.s$^{-1}$ at $f=600$ kHz in the case of the phase velocity, and from 0.05\,Np.m$^{-1}$ at 50 kHz to 0.1\,Np.m$^{-1}$ at 600 kHz in that of the attenuation.

\begin{figure}[htbp]
\centering
\begin{tabular}{cc}
(a) & (b)\\
\includegraphics[scale=0.44]{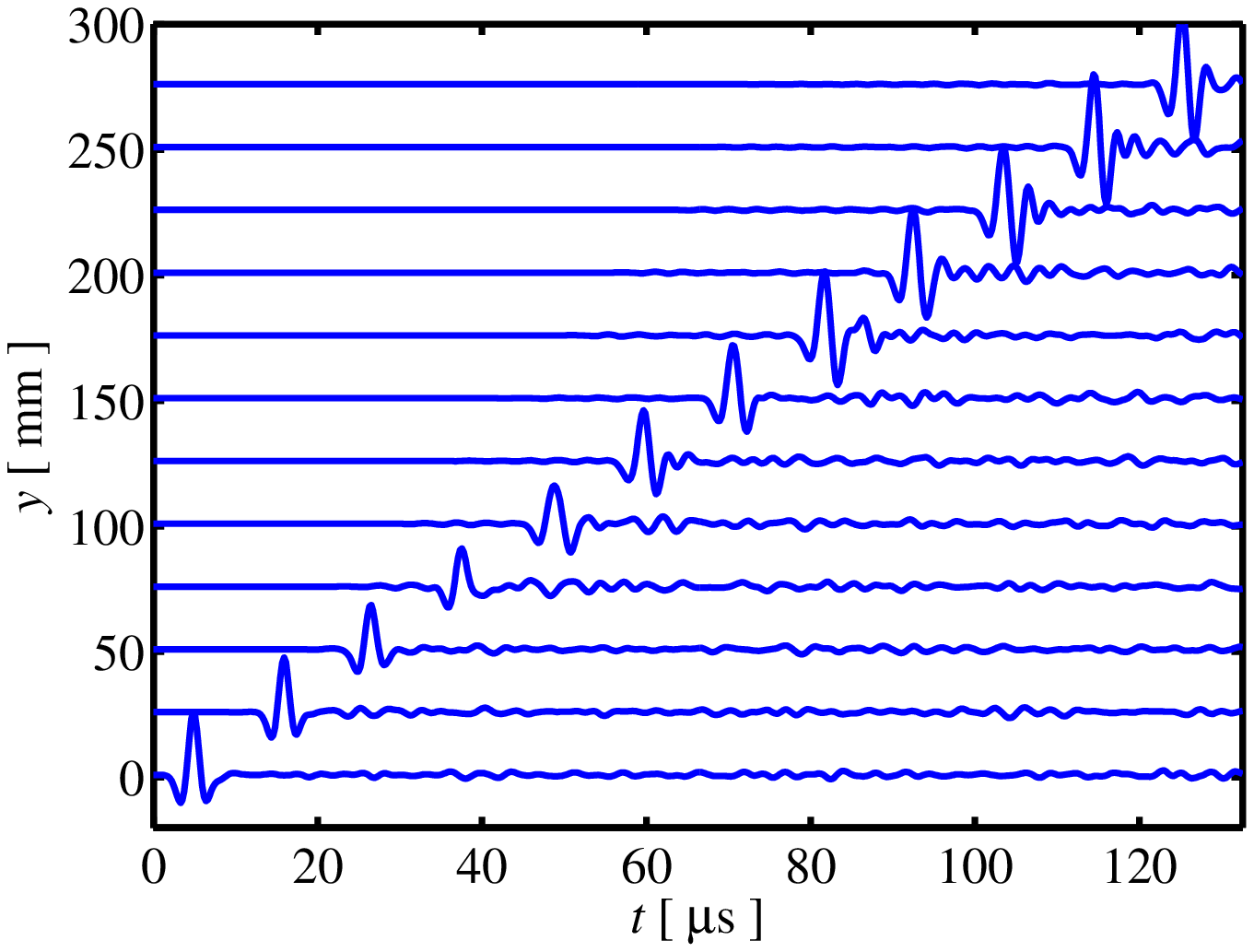}&
\includegraphics[scale=0.44]{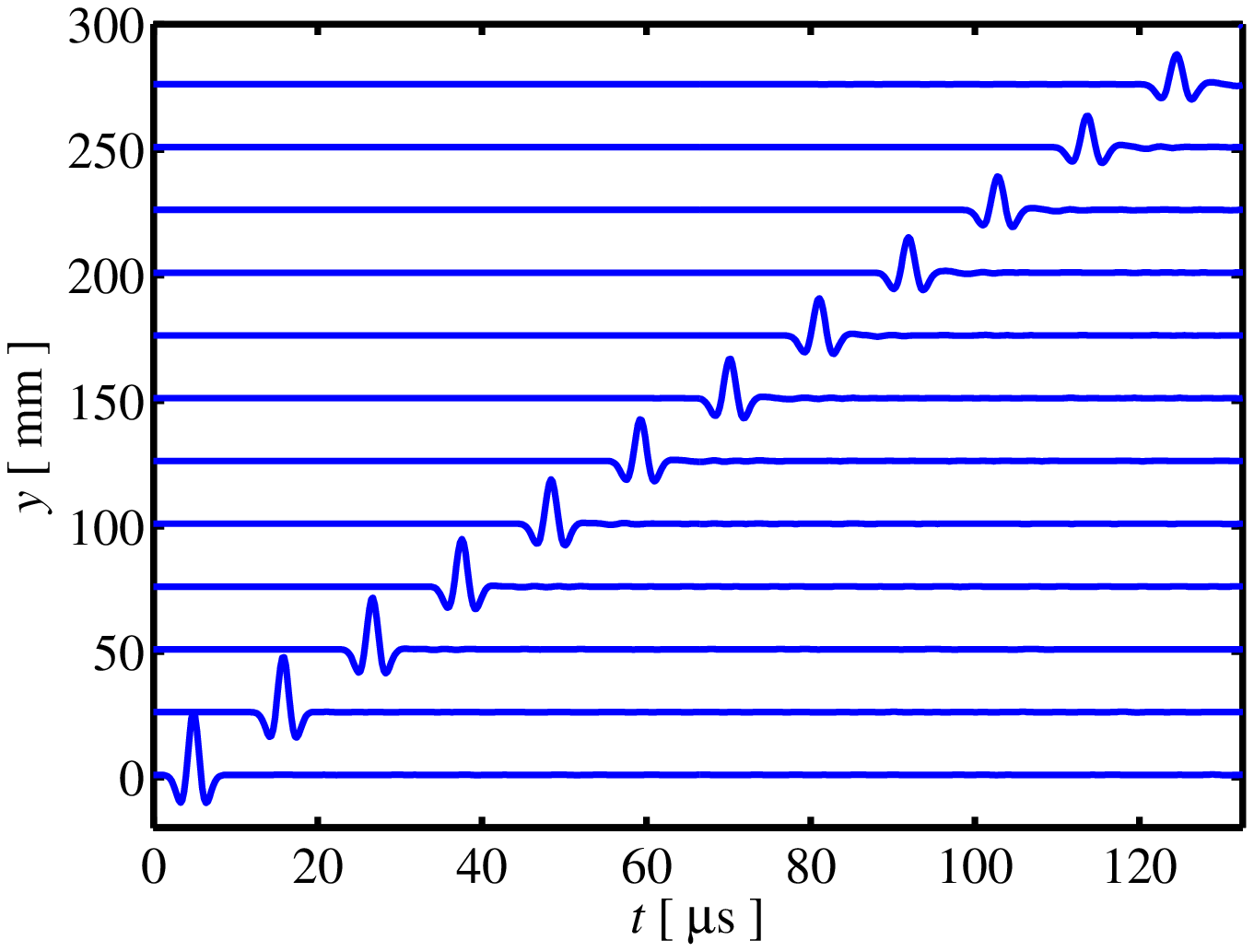}
\end{tabular}
\caption{Incident plane $T$-wave in a medium with 24\,\% inclusion concentration: (a) signals recorded along an array, (b) coherent signals obtained after summation.}
\label{FigChampIncCoh}
\end{figure}

Signals recorded along 2 different arrays logically show different behaviors. Figure \ref{FigChampIncCoh}-a gives the time histories at various receivers along one particular array. A main wave train is clearly visible in each of the time histories, followed by a coda. 

As recalled in section \ref{SecTDS1}, a coherent signal can be obtained by averaging the signals recorded on the various arrays \cite{Derode01}. The $v_{y}$ component is used in the case of an incident $L$-wave, whereas the $v_{x}$ component is used in that of an incident $T$-wave (stress components provide the same results). An example of coherent seismogram is presented in figure \ref{FigChampIncCoh}-b. The coda has disappeared, and the main wave train behaves like a plane wave propagating in a homogeneous (but dispersive and attenuating) medium. 
 
\subsection{Convergence of the coherent field to the effective field}\label{SecExpConverge}
 
The coherent field is obtained by averaging the signals recorded along the 300 arrays of receivers. If this number is too low, the estimated properties will still be dependent on the configuration of the scatterers encountered. Theoretically, the effective wavenumber can be defined by taking an infinite number of configurations, which means that all the possible configurations of scatterers will be taken into account; but this approach is obviously impracticable. The aim of this paragraph is to show that 300 arrays suffice for estimating the effective field and hence, the effective wavenumber.

\begin{figure}[htbp]
\centering
\includegraphics[width=10cm]{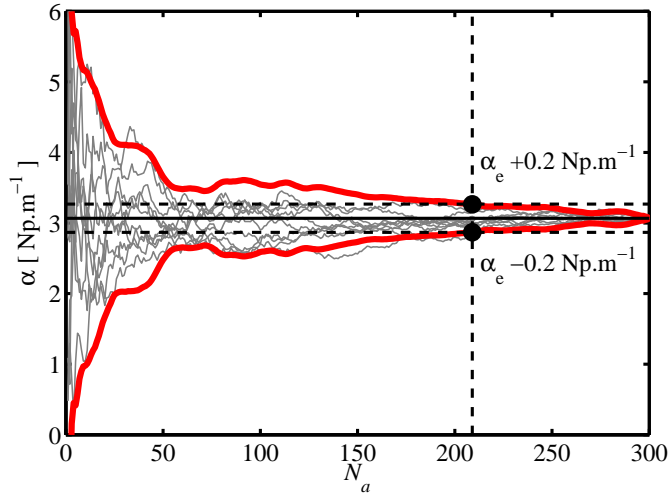}
\caption{Evolution of $\alpha_T$ in terms of $N_a$ at $f=300$\,kHz, $\phi=36$\,\%.}
\label{fig:exemple_C36S_alpha}
\end{figure}

In each case (in terms of the density $\phi$ of the scatterer and the incident wave $\beta=L,T$), a coherent signal is computed with an increasing number $N_a$ of arrays ranging from 1 to 300. The $N_a$ arrays are chosen randomly among the 300 available ones, to avoid taking consecutive arrays which are located too near each other in the medium. The properties $c_{\beta}$ and $\alpha_{\beta}$ are then evaluated from the averaged signal, and their evolution with $N_a$ is then studied at a given frequency. Figure \ref {fig:exemple_C36S_alpha} shows how $\alpha_T$ evolves with $N_a$, at $\phi=36$\% and $f=300$ kHz. As this evolution depends on the $N_a$ arrays selected in the averaging procedure, the study is repeated 10 times.

The value obtained by summing the fields over the 300 arrays is taken as a reference value for the effective medium: in this case, it was $\alpha_{e,T}$=3.06\,Np.m$^{-1}$. In figure \ref{fig:exemple_C36S_alpha}, the red curve gives the envelope of all 10 curves in gray, corresponding to 2 standard deviations of $\alpha_T(N_a)$. As was to be expected, as $N_a$ increases, the value of $\alpha_T$ tends towards the reference value $\alpha_{e,T}$. This figure shows that when there are too few configurations, $\alpha_{e,T}$ cannot be accurately assessed; for instance, taking only 50 configurations results in an uncertainty greater than 1\,Np.m$^{-1}$. Taking $\pm$0.2\,Np.m$^{-1}$ to be an acceptable level of uncertainty for $\alpha_{e,T}$, the optimum number $N_{opt}$ of configurations requested must be greater than $N_{opt}\simeq 210$ in the case of the present example. 

\begin{figure}[htbp]
\centering%
\begin{tabular}{cc}
 effective phase velocity ($L$ wave) & effective phase velocity ($T$ wave)\\
\includegraphics[width=6cm]{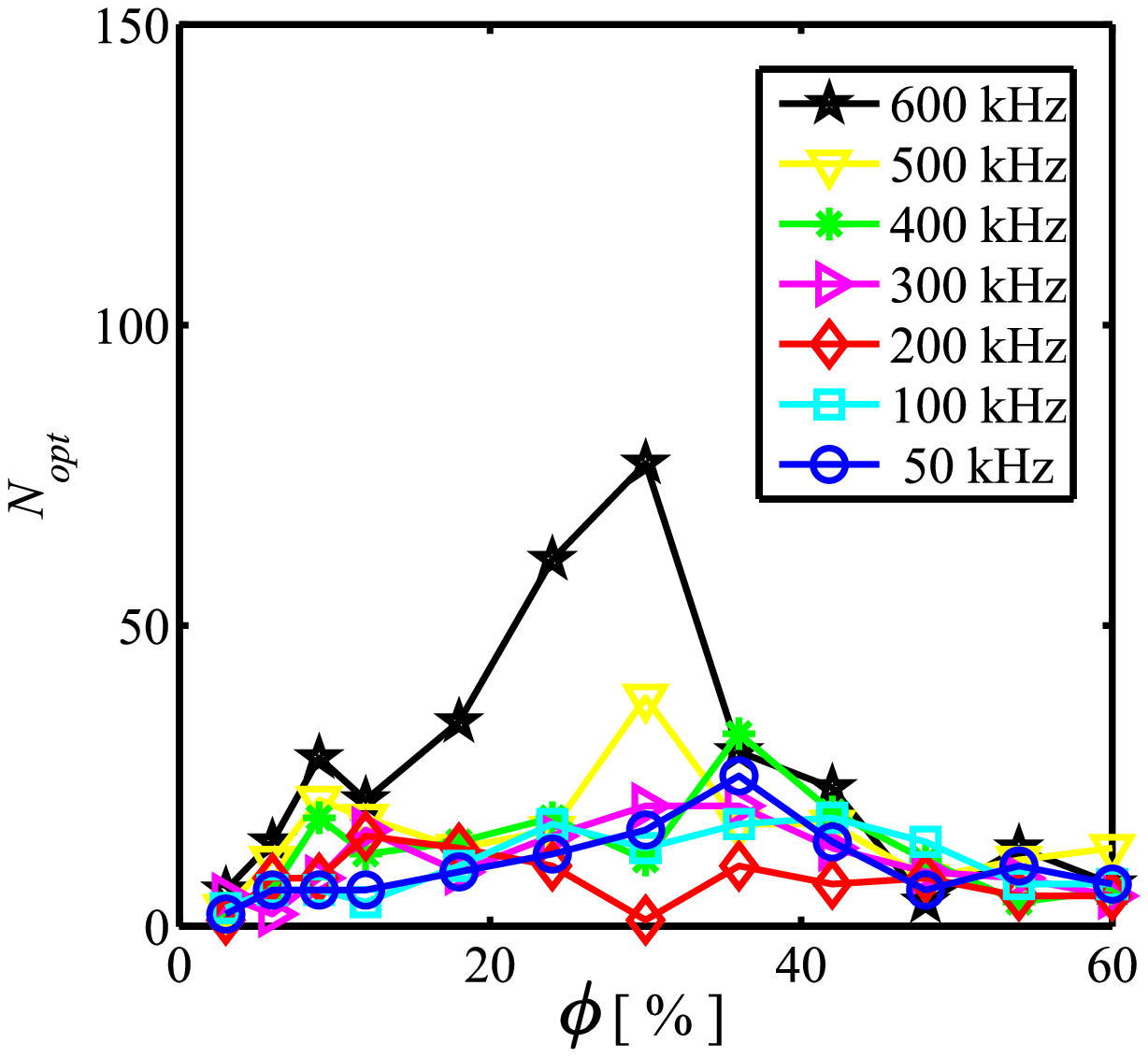}&
\includegraphics[width=6cm]{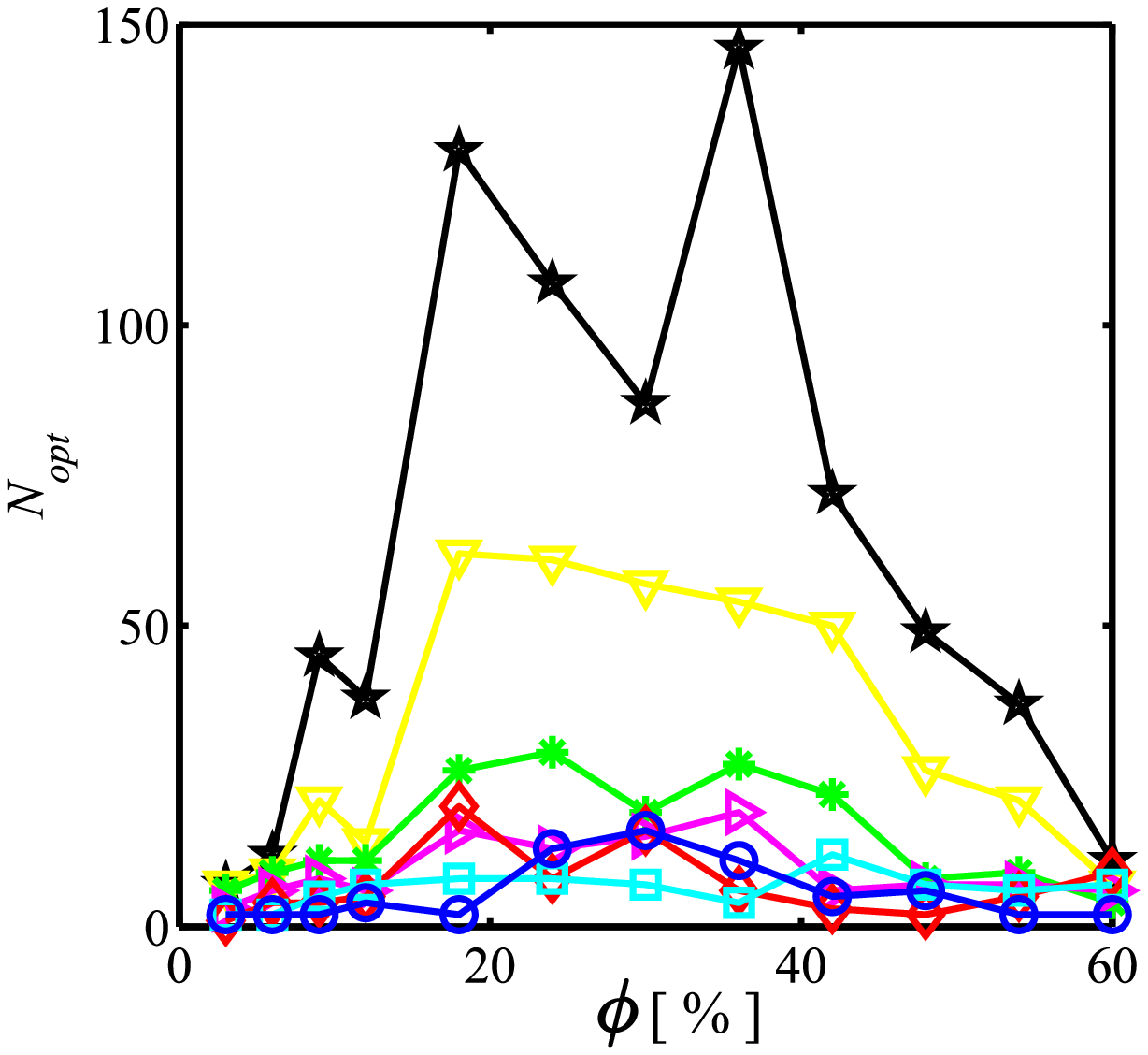}\\
effective attenuation ($L$ wave) & effective attenuation ($T$ wave)\\
\includegraphics[width=6cm]{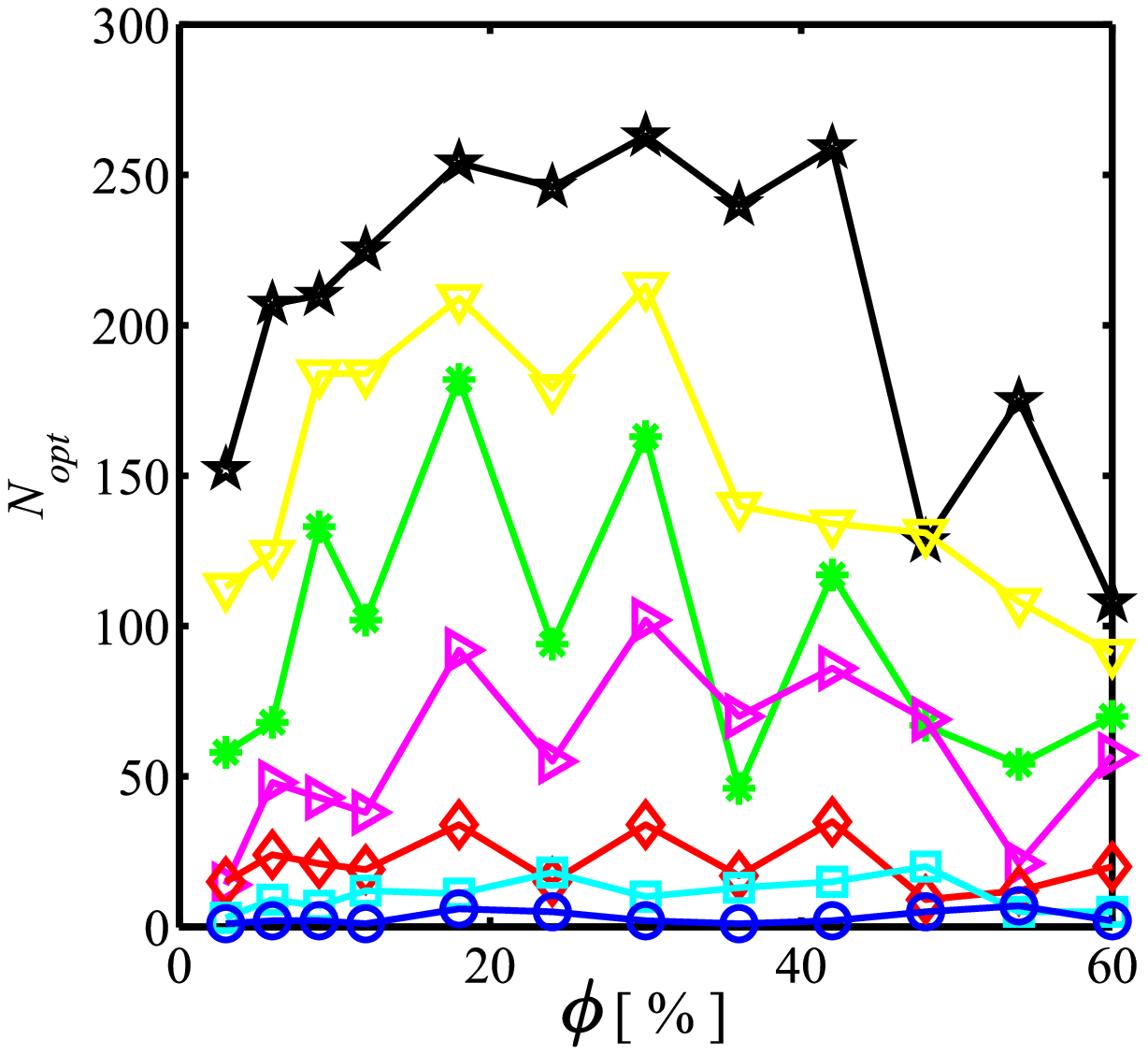}&
\includegraphics[width=6cm]{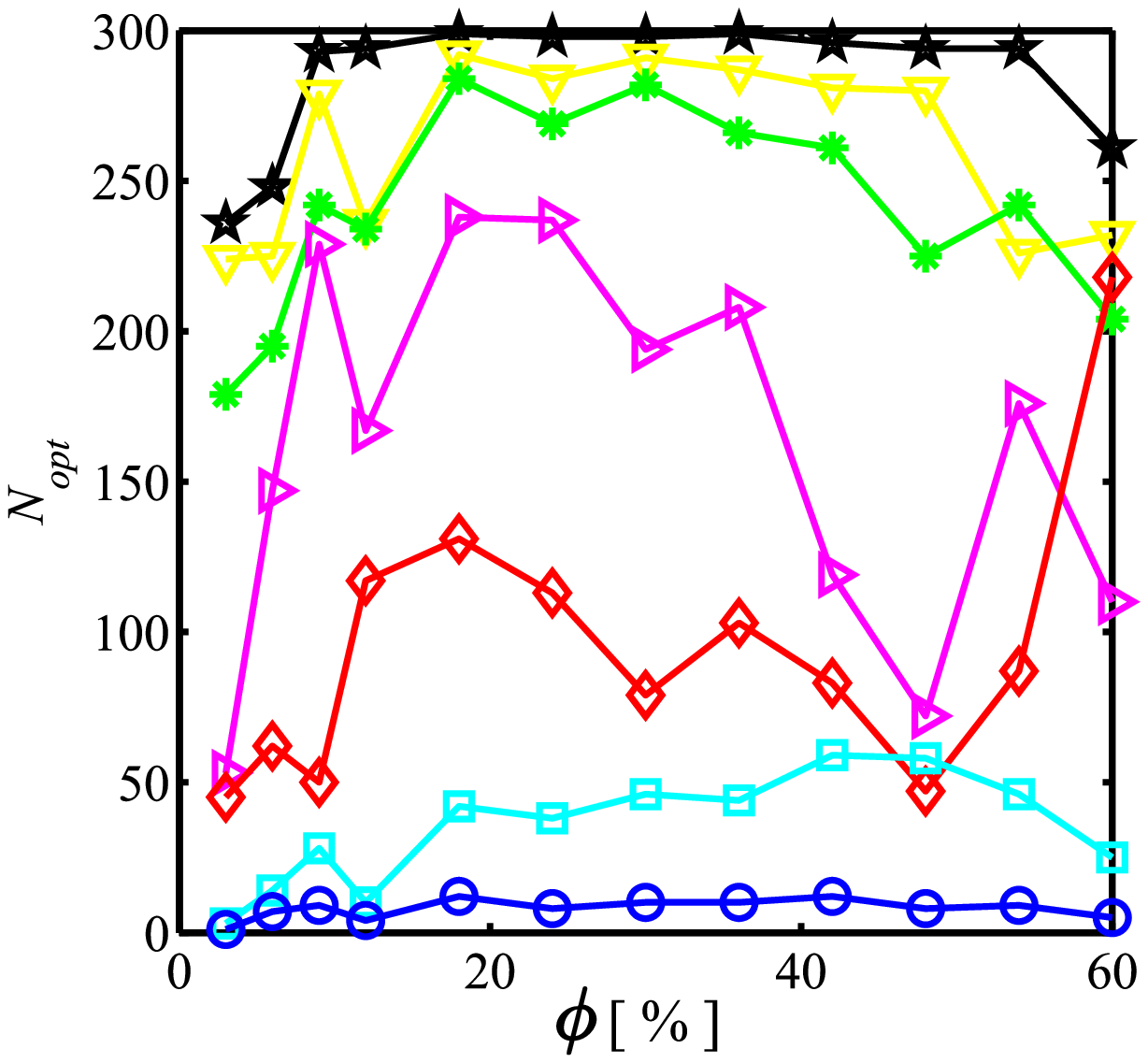}
\end{tabular}
\caption{Optimum number $N_{opt}$ of configuration of scatterers for determining $c_{e,\beta}$ or $\alpha_{e,\beta}$ at all densities.}
\label{fig:conv_resultats}
\end{figure}

Figure \ref{fig:conv_resultats} summarizes the values of $N_{opt}$ obtained in all the cases studied, at several frequencies, using the procedure above. The criterion used to obtain an accuracy of about $c_{e,\beta}$ was $\pm$5\,m.s$^{-1}$; in the case of $\alpha_{e,\beta}$, we took $\pm$0.2\,Np.m$^{-1}$. Except for the higher frequencies ($>$ 400\,kHz), only about 20 configurations are required for assessing the effective velocity phase $c_{e,\beta}$, whereas greater values of $N_{opt}$ are required for assessing the effective attenuation. If $N_{opt}$ is obtained at a given scatterer concentration, its value will increase with the frequency at a given polarity, and it will be almost twice as high with $T$-waves as with $L$-waves at a given frequency. Convergence therefore depends mainly on the size of the wavelength, as the effects of multiple scattering are greater at shorter wavelengths.

The results obtained on $\alpha_T$ at high frequencies (500 and 600\,kHz) show that the optimum number of configurations $N_{opt}$ was almost 300, which was the maximum number of configurations available, whatever the density of the scatterers. In the present 2 cases, it was not possible to say whether the optimum number of configurations was actually reached, so as to be able to assess the attenuation with a sufficiently high level of certainty. In all the other cases, the $c_{e,\beta}$ and $\alpha_{e,\beta}$ values obtained based on 300 scatterer configurations were fully representative of the effective wavenumber.


\section{Numerical results}\label{SecRes}

\subsection{Simulated effective wavenumbers}\label{SecResWN}

\begin{figure}[htbp]
\begin{center}
\begin{tabular}{cc}
effective phase velocity ($L$ wave) & effective phase velocity ($T$ wave) \\
\includegraphics[width=6.6cm]{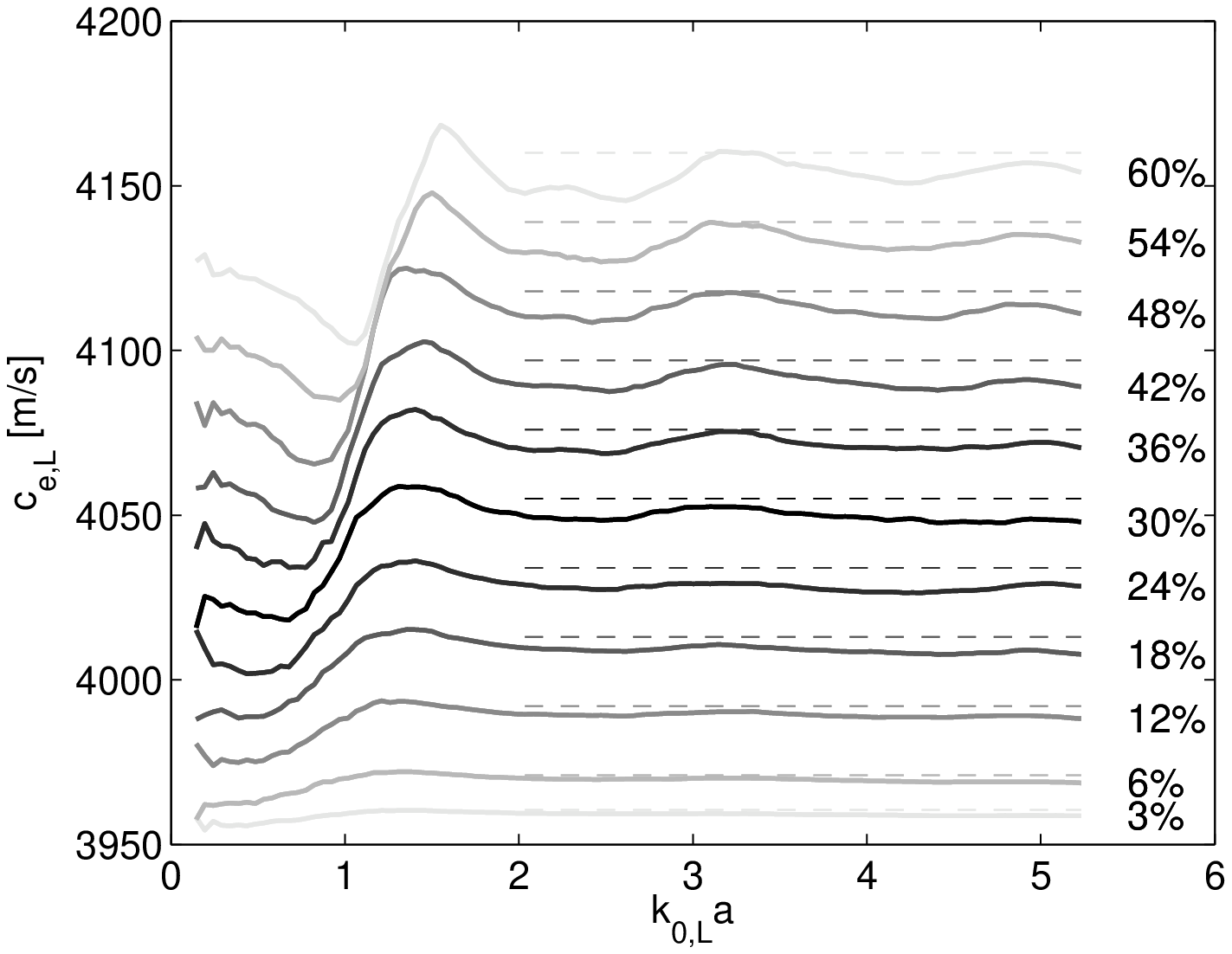} &
\includegraphics[width=6.6cm]{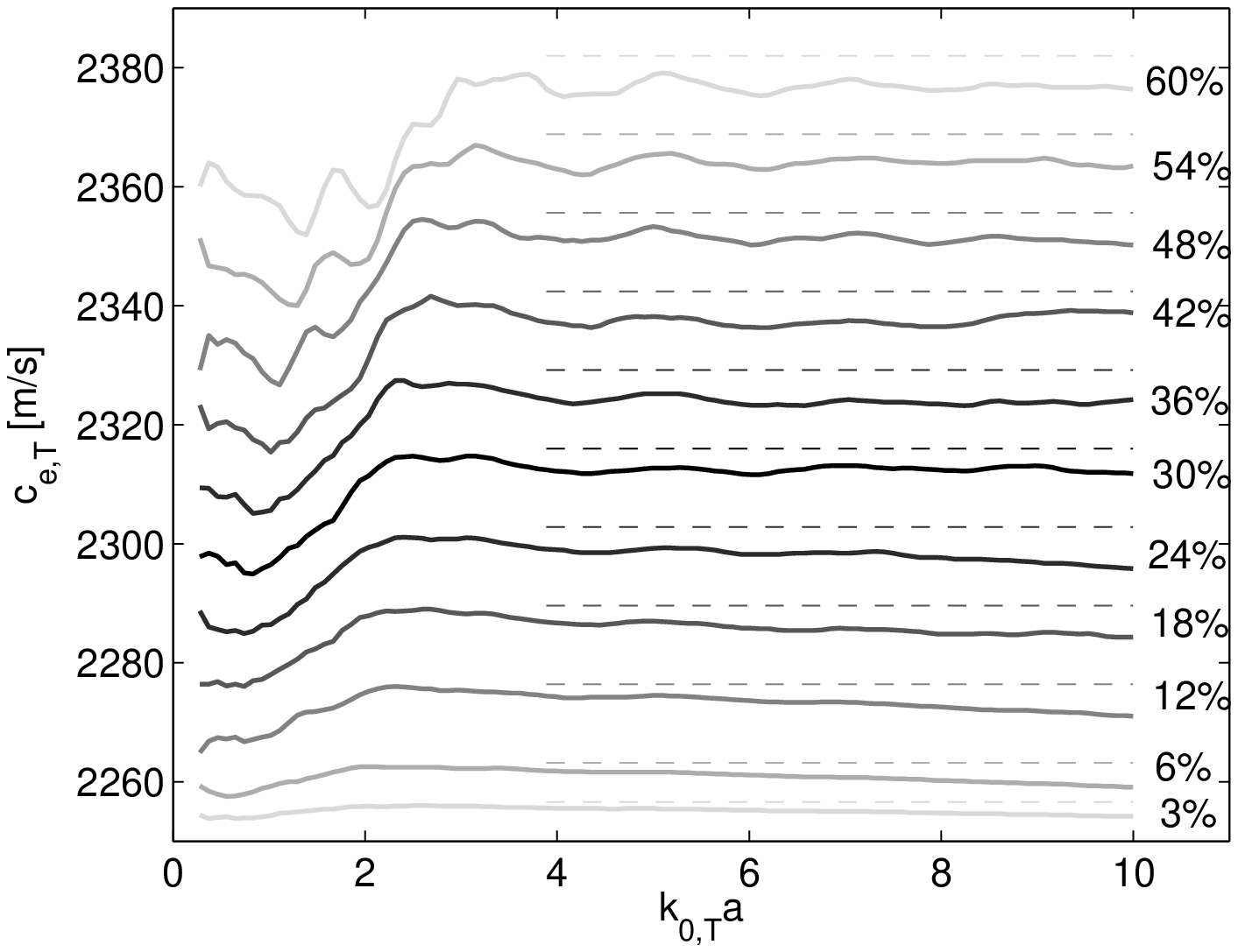}
\\
effective attenuation ($L$ wave) & effective attenuation ($T$ wave)\\
\includegraphics[width=6.6cm]{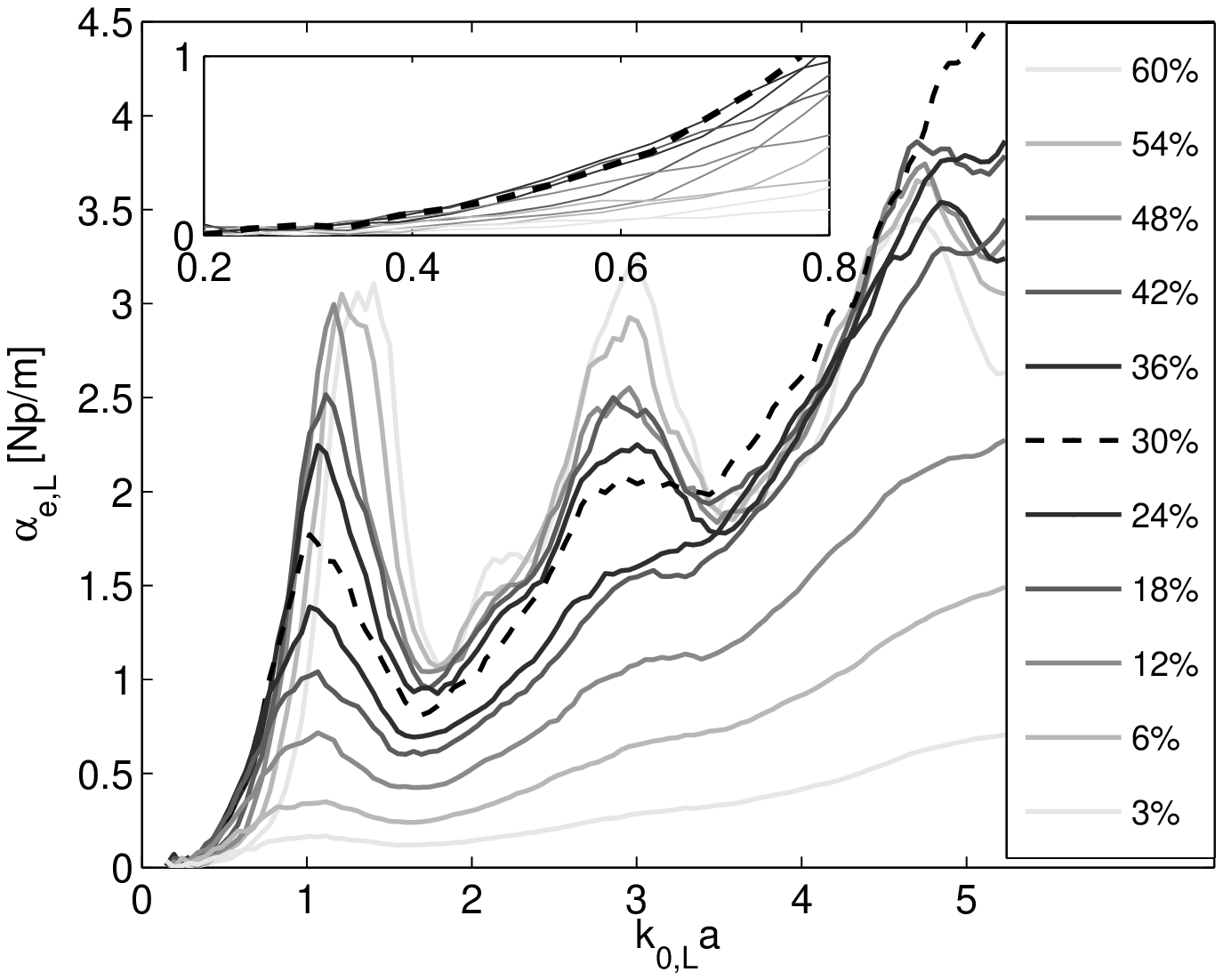} &
\includegraphics[width=6.6cm]{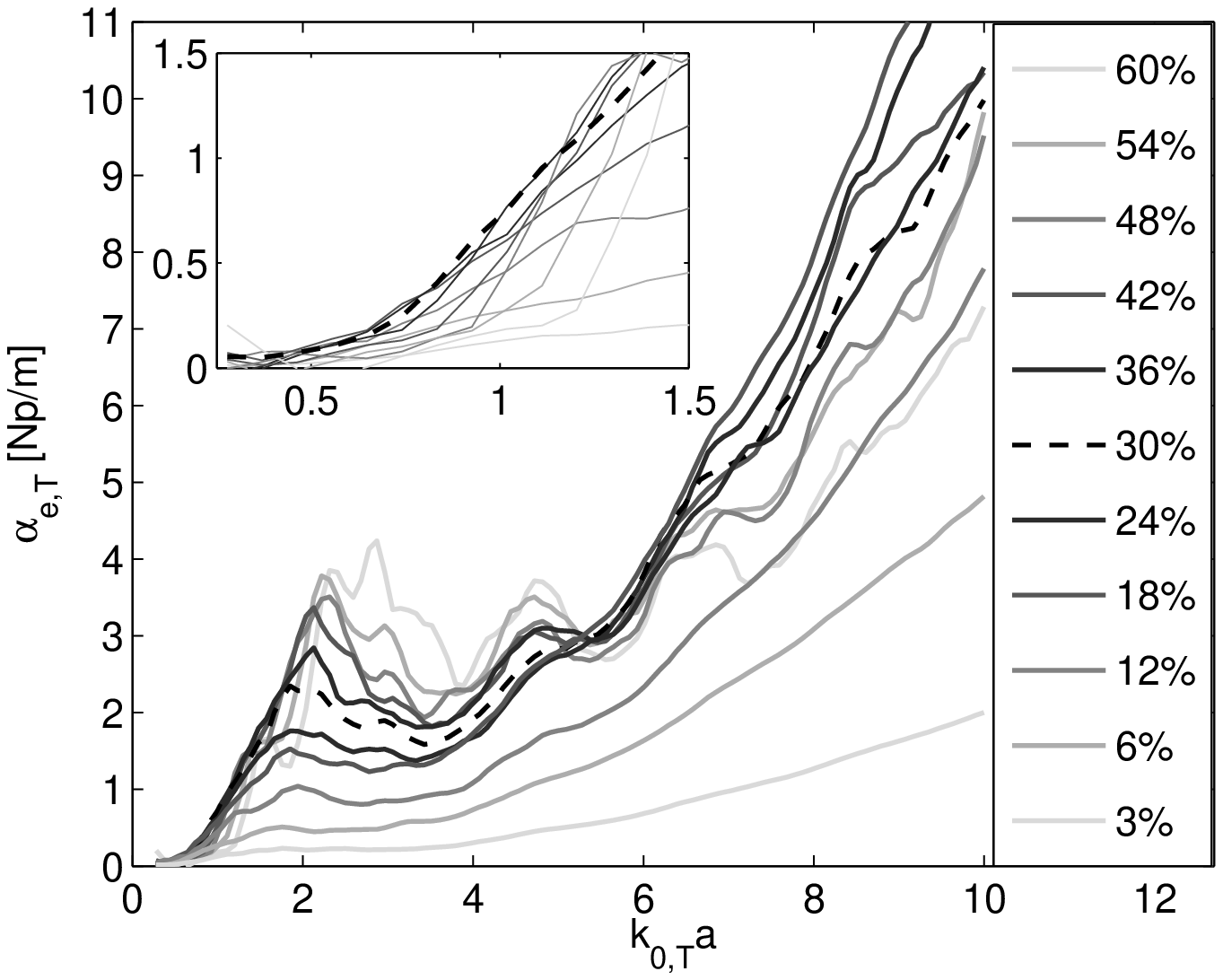}
\end{tabular}
\end{center}
\caption{Effective properties obtained with numerical simulations and signal processing tools, with  longitudinal (left) and transverse (right) incident waves at various inclusion concentrations $\phi$. Top: phase velocity $c_{e,\beta}$; the horizontal dashed lines give the mean phase velocity $\bar{c}$. Bottom: attenuation $\alpha_{e,\beta}$ where $\phi=30\%$ (that is, around the critical threshold mentioned in the discussion) is given by a dashed line; the insert is a zoom in the low-frequency range.}
\label{fig:SimulVphiAtt}
\end{figure}

Results obtained in the numerical simulations with the two polarizations of the incident wave ($\beta=L,\,T$) and at various scatterer concentrations $\phi$ are presented in figure \ref{fig:SimulVphiAtt}. In a first approximation, the effective phase velocity was found to be proportional to the density of the inclusions, increasing monotonically with $\phi$. The effective phase velocity also showed a dispersive behavior, which became more conspicuous as $\phi$ increased. This effect was stronger at the lower frequencies (at $k_{0,L}a\lesssim 1$ and $k_{0,T}a \lesssim 2$), where the value of the overall phase velocity was lower than at high frequencies. A maximum value of $c_e$ was reached at $k_{0,L}a \approx 1$ and $k_{0,T}a \approx 2$. At higher frequencies, $c_{e,\beta}$ remained almost constant but small fluctuations are visible, up to 10\,m.s$^{-1}$ at 60\% which amount to less than 0.5\% of the phase velocity. However, the positions of these local extrema are almost $\phi$-invariant. At high frequencies, the mean phase velocity $\bar{c}=(1-\phi)\,c_{0,\beta}+\phi\,c_{1,\beta}$ is a good approximation in the case of dilute media and an upper limit in that of more densely packed media.

The attenuation is more difficult to explain: contrary to what occurs with the phase velocity, the attenuation does not depend monotonically on the concentration. The frequency dependence shows $\phi$-invariant peaks corresponding to the local maxima of the phase velocity, mainly at $k_{0,L}a \approx 1$ and $k_{0,L}a \approx 3$ in the case of $L$-waves, and $k_{0,T}a\approx 2$ and $k_{0,T}a\approx 5$ in that of $T$ waves. In dilute media ($\phi \lesssim 20\%$), the attenuation is proportional to $\phi$. In denser heterogeneous media, the behavior depends on both the inclusion concentration and the frequency range. Three types of overall behaviors were observed: 
\begin{itemize}
\item around the previously mentioned peaks, $\alpha_{e,\beta}$ increases with $\phi$; 
\item at high frequencies between these peaks, $\alpha_{e,\beta}$ remains at an almost constant value when $30\% \lesssim \phi \lesssim 60\%$; 
\item at low frequencies, $\alpha_{e,\beta}$ reaches a maximum at $\phi \approx 30\%$, and then it decreases.
\end{itemize}
With an incident $T$-wave and $k_{0,T}\,a>7$, the behavior is less clear-cut: the attenuation reaches a peak at around $\phi\simeq 30\%$. However, the accuracy of these findings was not confirmed in the section \ref{SecExpConverge}.


\begin{figure}[htbp]
\begin{center}
\begin{tabular}{c}
phase velocity: ISA, WT, CN\\
\includegraphics[width=8.125cm]{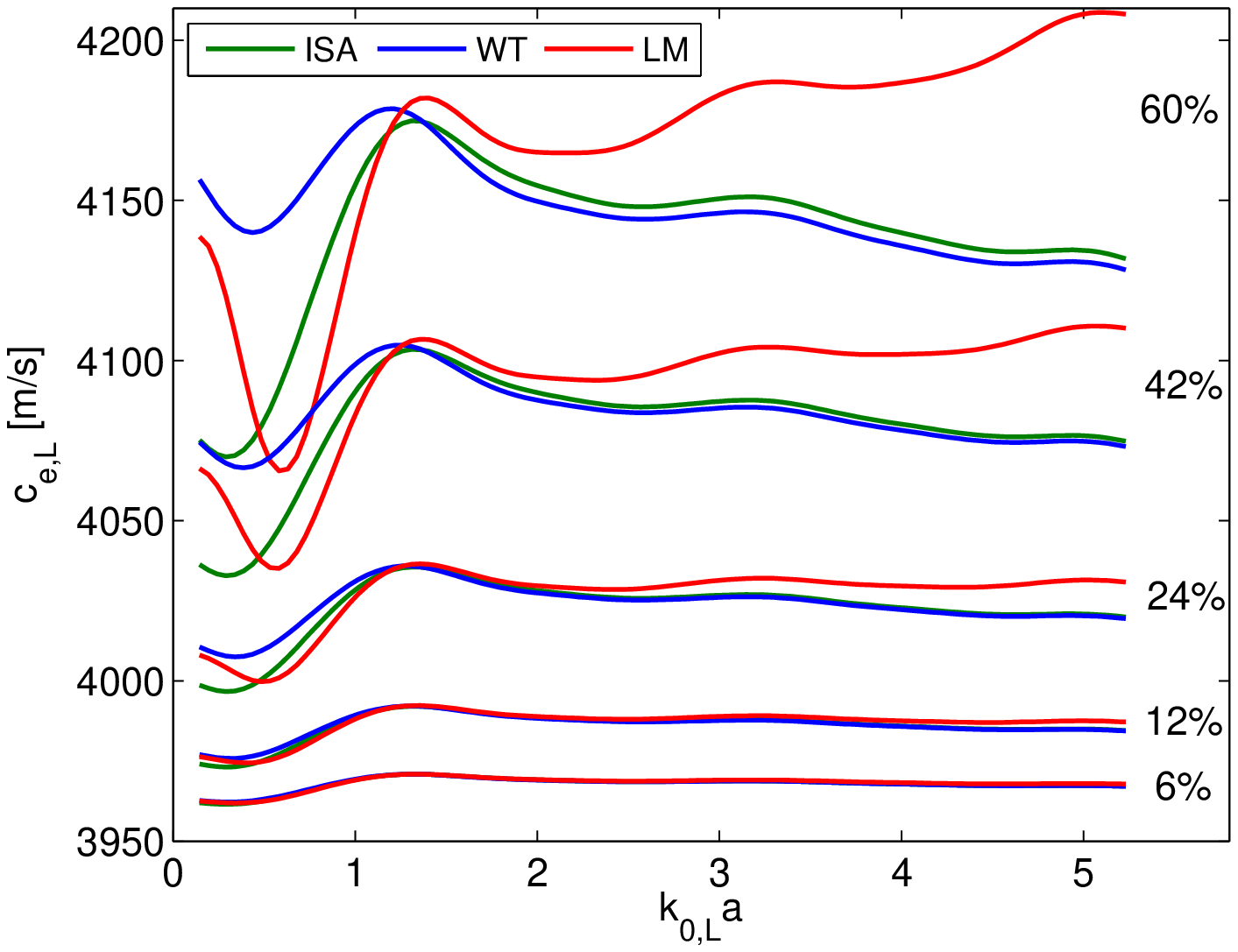}\\
\\
attenuation: ISA, WT\\
\includegraphics[width=8.125cm]{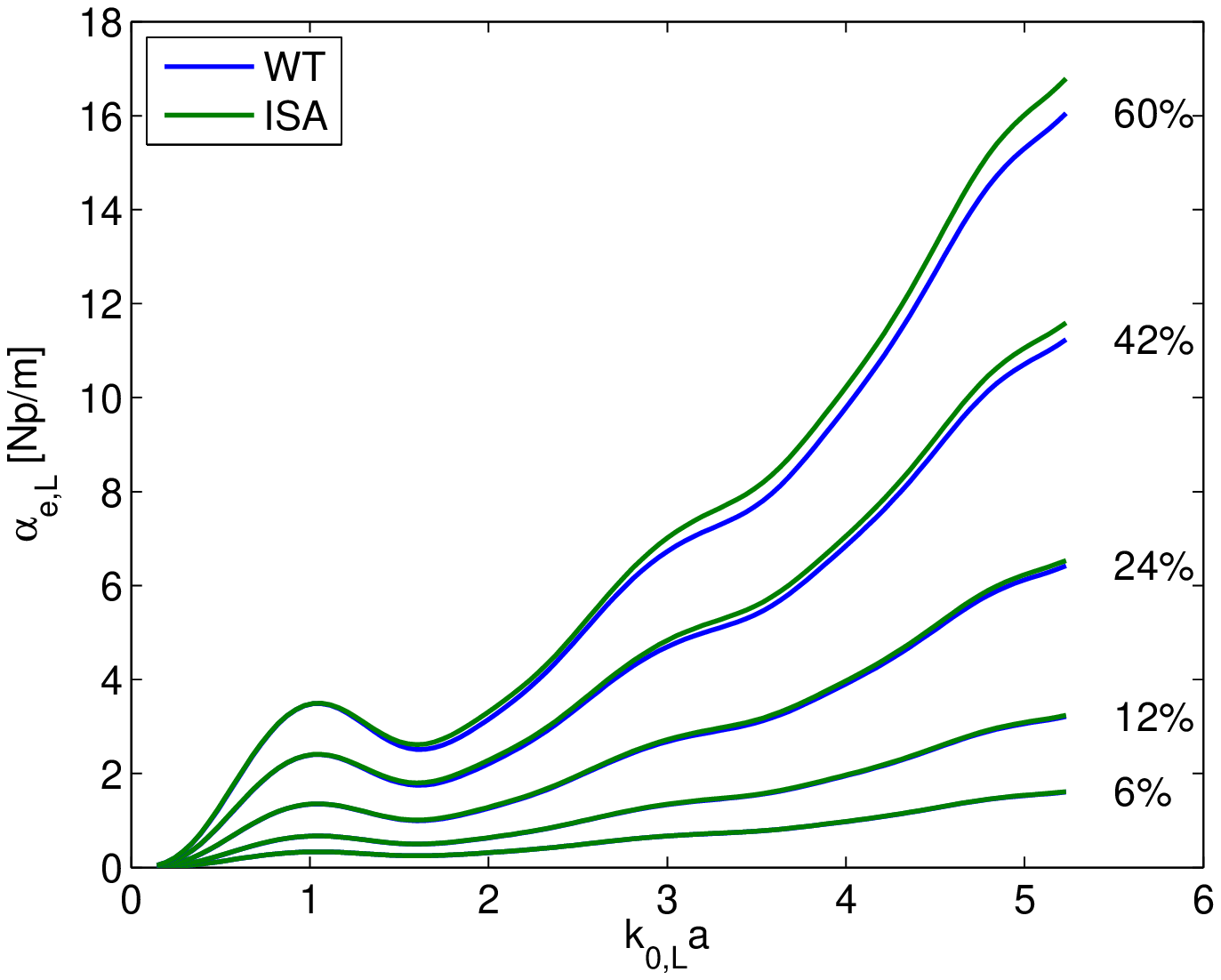}\\
\\
attenuation: CN\\
\includegraphics[width=8.125cm]{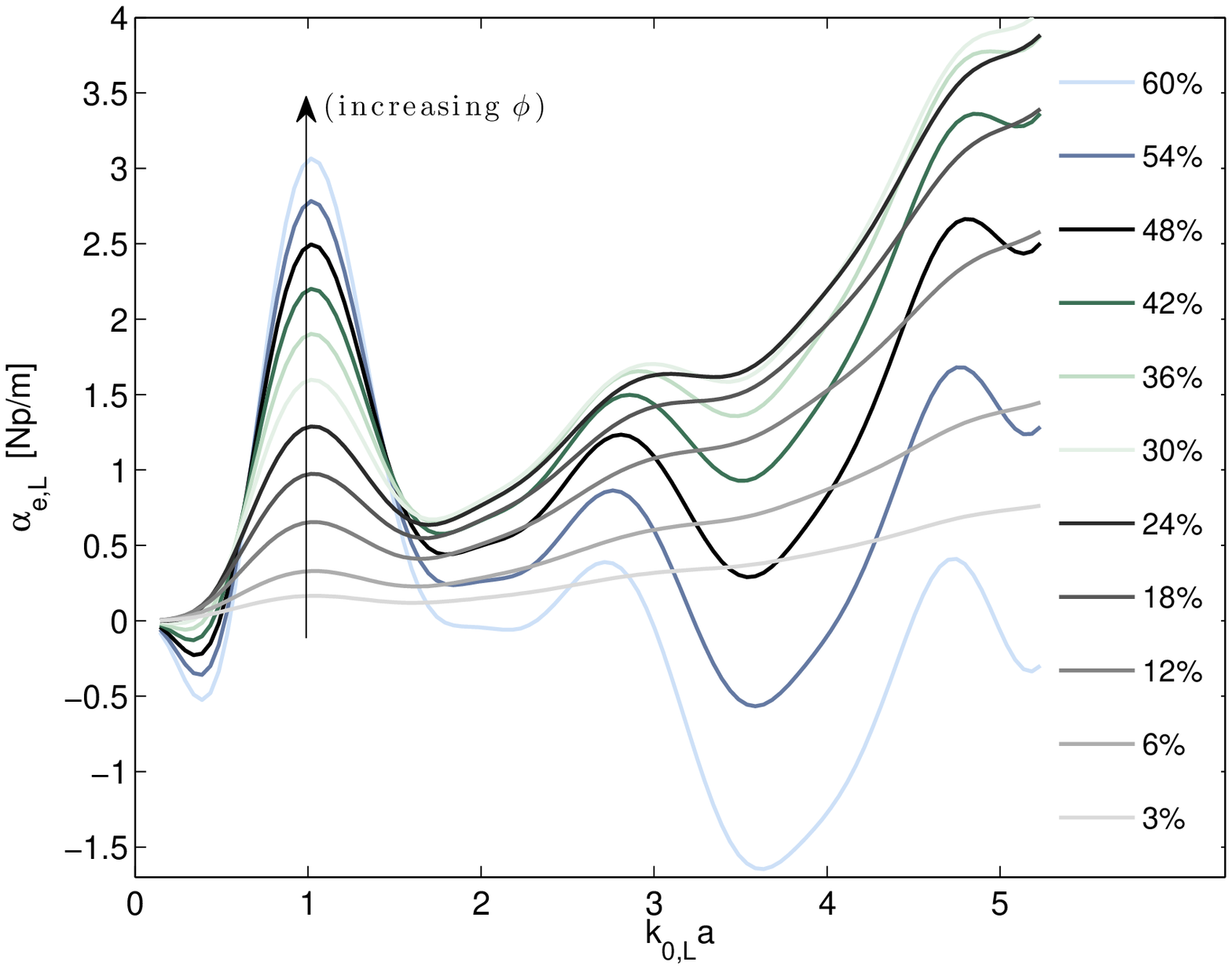}
\end{tabular}
\end{center}
\caption{Effective properties of the longitudinal incident waves obtained with the Independent Scattering Approximation (ISA), Waterman-Truell (WT) and Conoir-Norris (CN)} models.
\label{CompISAWT}
\end{figure}

\subsection{Comparison between various theoretical models}\label{SecResModel}

In this section, the effective wavenumbers predicted by the Independent Scattering Approximation (ISA), Waterman-Truell (WT) \cite{WatermanTruell61}, and recent Conoir-Norris (CN) \cite{Conoir10} multiple-scattering models are presented. Technical details can be found in the appendix \ref{AppEff}. The results obtained with the explicit (or analytical) formulation of these models are compared in figure \ref{CompISAWT} in the case of a longitudinal incident wave, and the same comments apply in the case of transverse waves.

ISA and WT models give similar results. The phase velocities differ only at low frequencies in the case of densely packed media, where both models are inaccurate; we will therefore focus on the WT predictions. As regards the phase velocity, all these models predicted the same overall behavior as the simulations: a monotonic increase with the concentration, the same $\phi$-invariant position of the local extrema, and an increase in the dispersion with the concentration. However, the attenuation given by the ISA and WT models are mainly linear functions of the concentration, contrary to what observed with the simulations: the attenuation occurring in densely packed media was clearly over-estimated. The attenuation predicted by the CN model differed considerably from that obtained with the previous models. First, the attenuation showed a linear dependence on the concentration only at low densities $\phi\lesssim 30\%$, where it reached a maximum (except for the peak at $k_{0,L}a \simeq 1$). Secondly, the attenuation was negative in some (low or high) frequency ranges, which is unphysical, but this occurs only at very high concentrations. Similar behavior has already be noticed \cite{Varadan85,Caleap12}.


\subsection{Comparison between theoretical models and numerical results}\label{SecResComparison}

\begin{figure}[htbp]
\begin{center}
\begin{tabular}{cc}
phase velocity: $L$ wave & phase velocity: $T$ wave\\
\includegraphics[width=6.6cm]{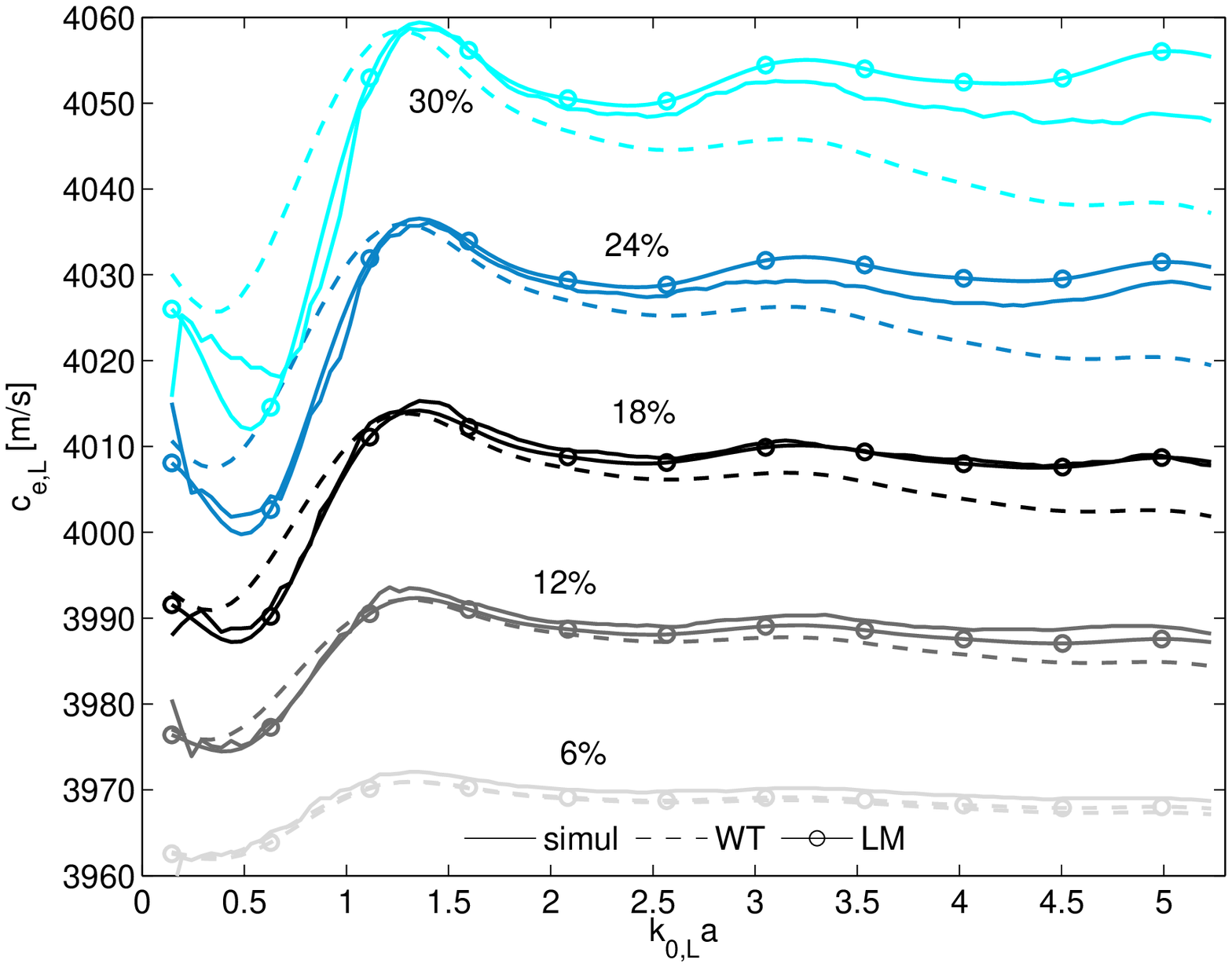}&
\includegraphics[width=6.6cm]{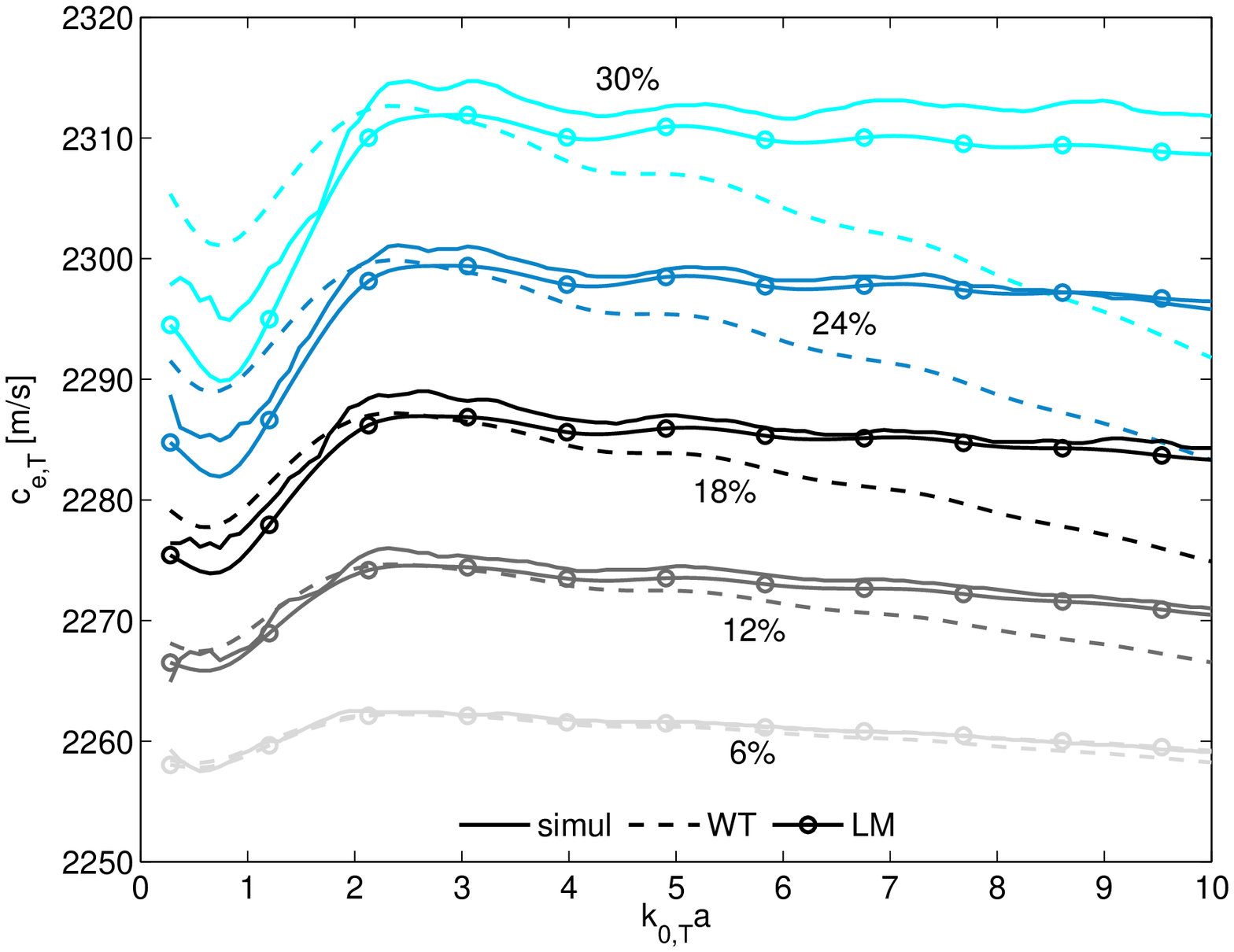}
\end{tabular}
\end{center}
\caption{Comparison between numerical simulations and modeling predictions of the phase velocity in the case of longitudinal (left) and transverse (right) incident waves at the lower inclusion concentrations ($\phi\leq30$\%).}
\label{compWTLMsimul_VIT}
\end{figure}

\begin{figure}[htbp]
\begin{center}
\begin{tabular}{cc}
attenuation: $L$ wave & attenuation: $T$ wave\\
\includegraphics[width=6.6cm]{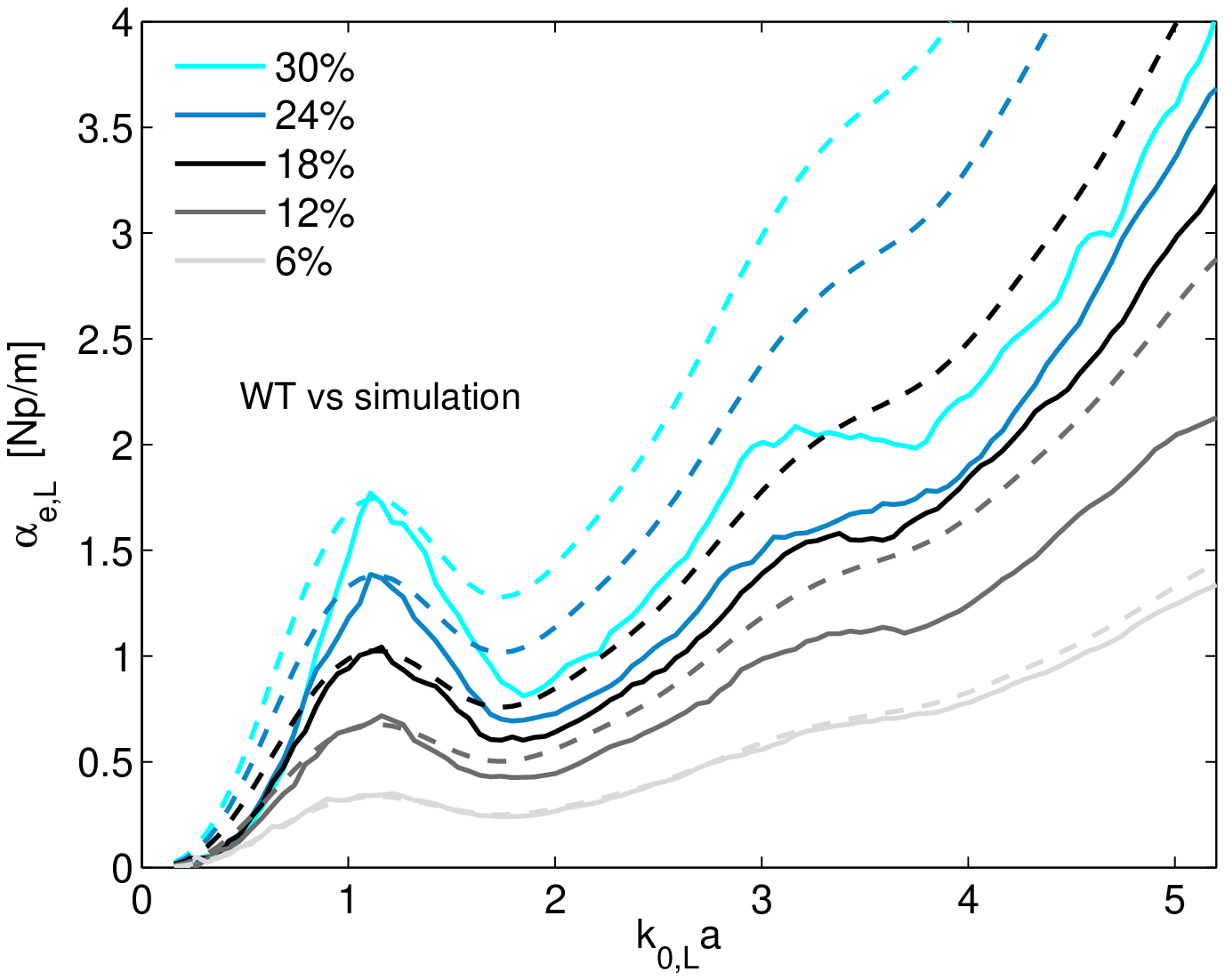}&
\includegraphics[width=6.6cm]{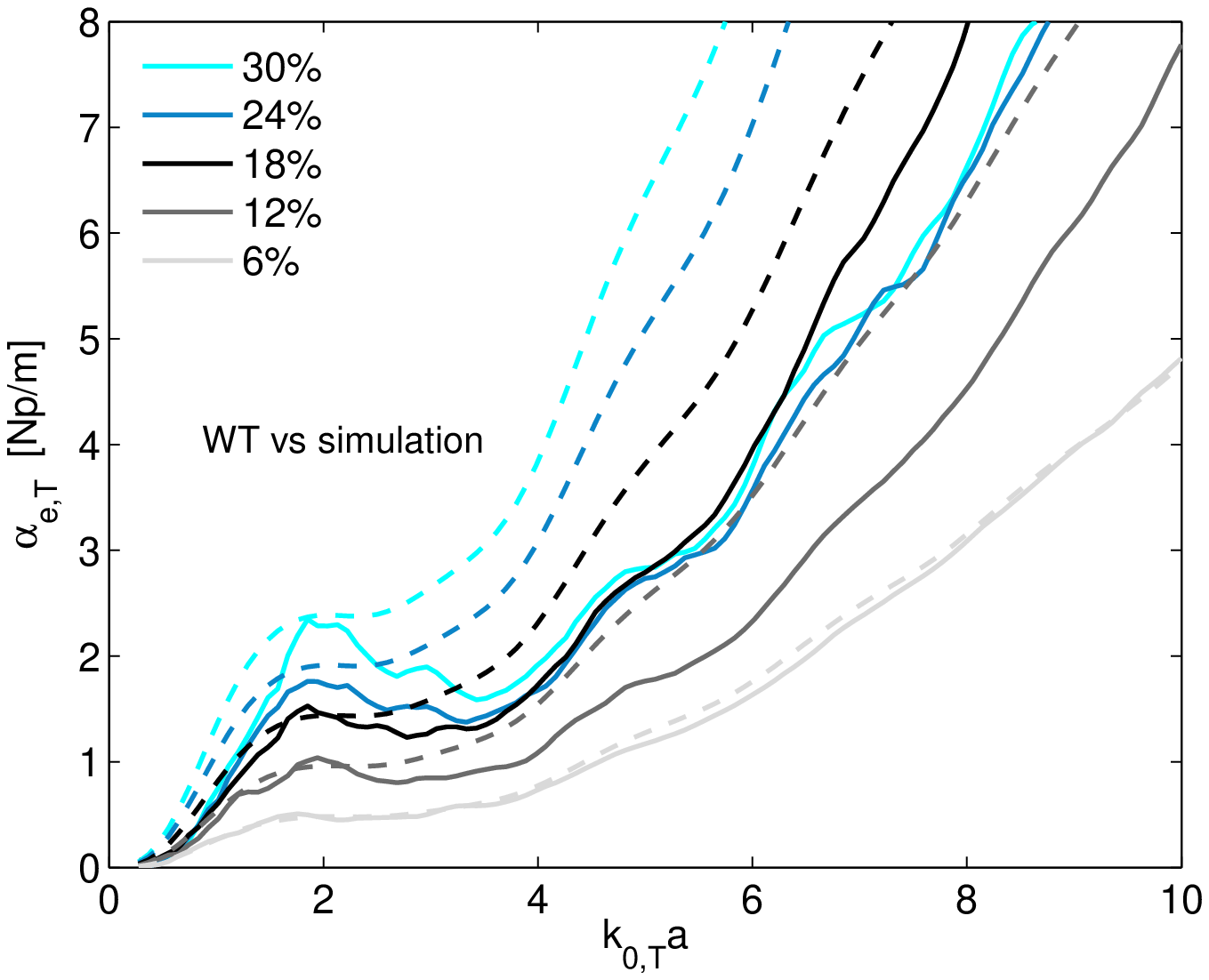}\\
\includegraphics[width=6.6cm]{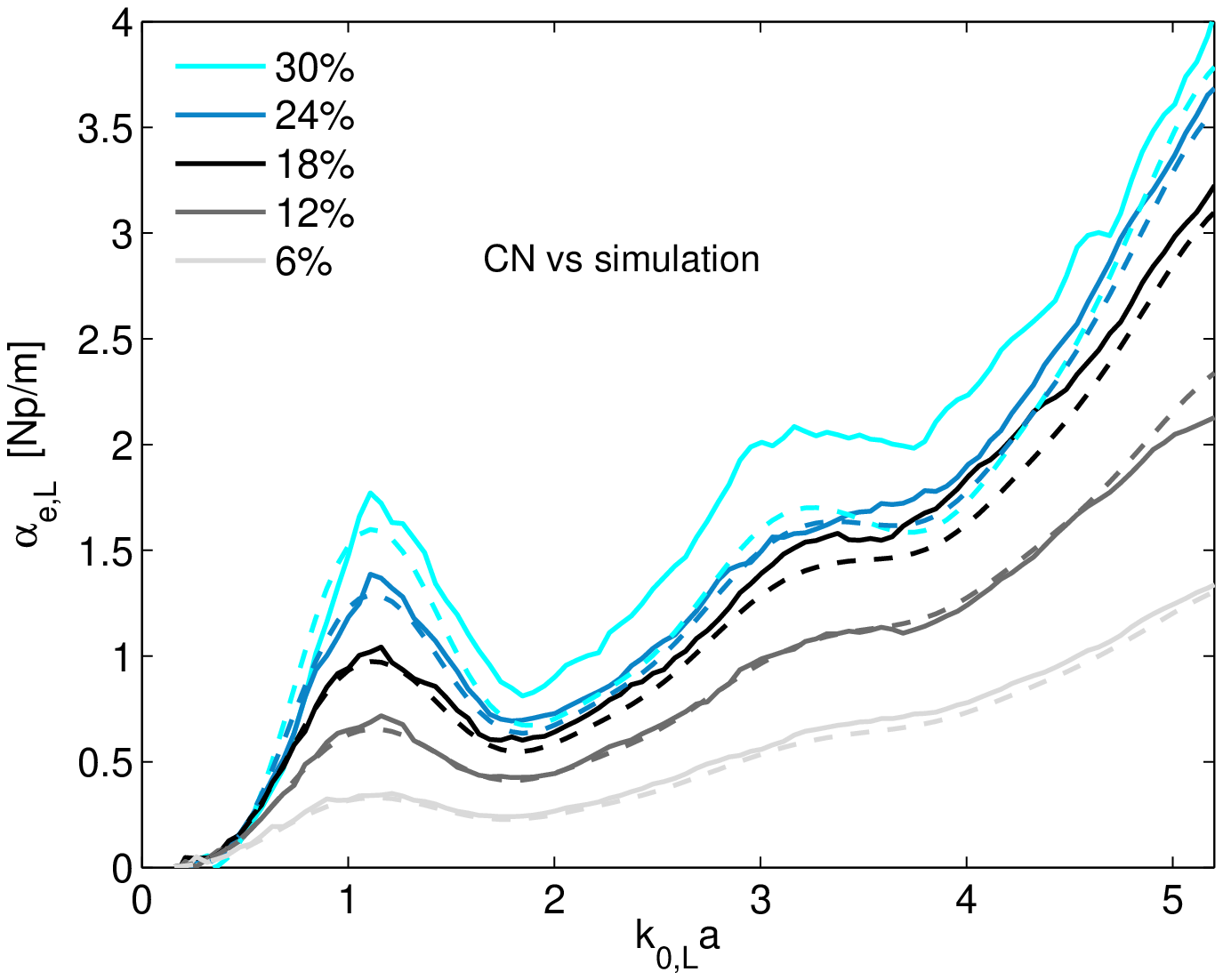}&
\includegraphics[width=6.6cm]{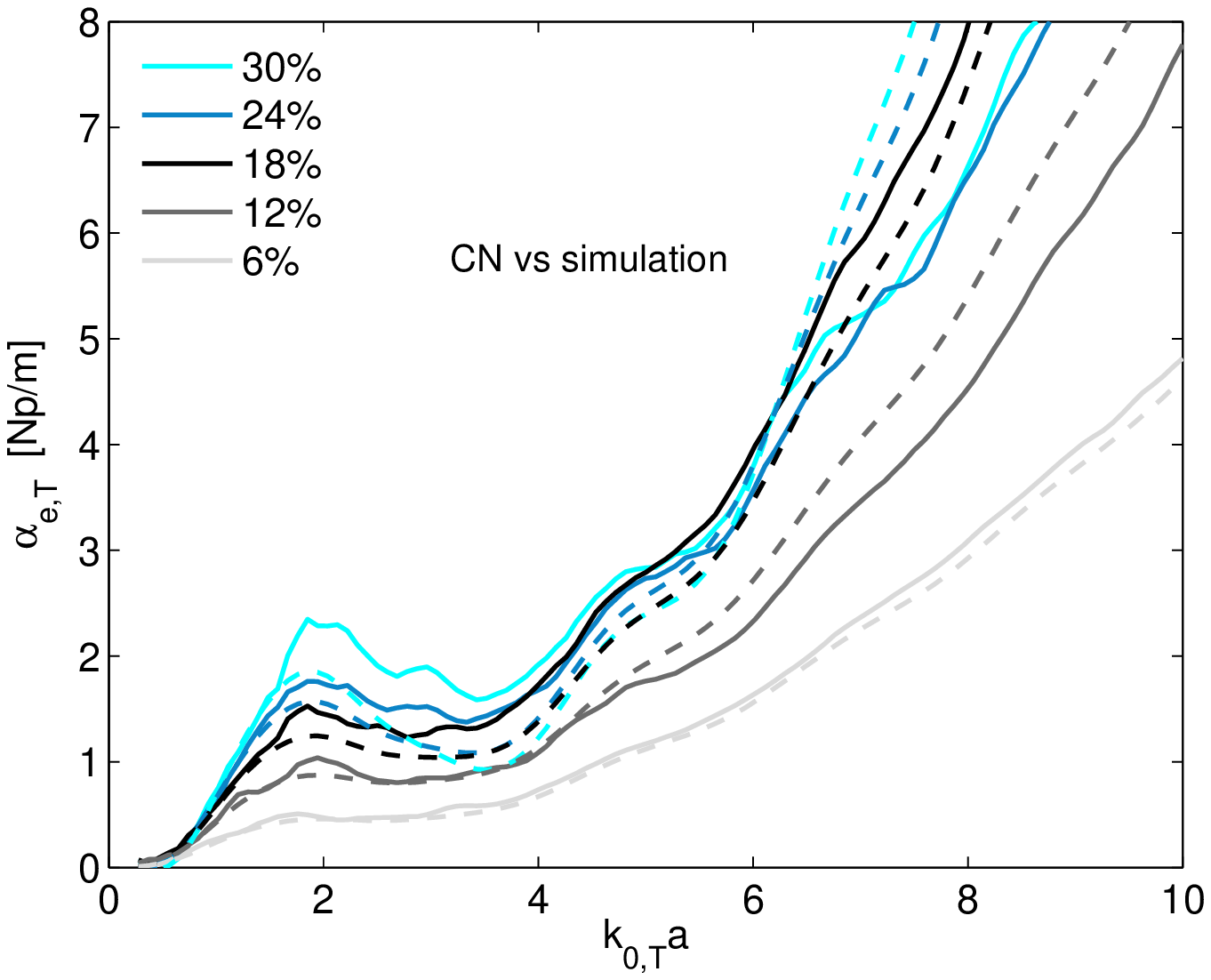}\\
\end{tabular}
\end{center}
\caption{Comparison between the effective attenuation obtained in numerical simulation and modeling predictions at the lower inclusion concentrations ($\phi\leq30$\%). Results obtained in the case of longitudinal (left) and transverse (right) incident waves versus the Waterman-Truell (top) and Conoir-Norris} predictions (bottom). Solid line: results of numerical simulations; dashed line: modeling predictions.
\label{compWTLMsimul_ATT}
\end{figure}

These comparisons between models and simulations made it possible to determine the range of validity of each model (see figure \ref{compWTLMsimul_VIT} for the phase velocity and figure \ref{compWTLMsimul_ATT} for the attenuation). In view of the signal processing limitations, satisfactorily results were defined as those with an error of less than 5\,m.s$^{-1}$ in the case of the phase velocity, and less than 0.2\,Np.m$^{-1}$ in that of the attenuation. On this basis, the WT model is suitable for dealing only with very dilute media $\phi\lesssim 12\%$ for $c_e$, and $\phi \leq6\%$ for $\alpha_e$, whereas the CN model gives accurate results up to inclusion concentrations of $24\%$ in terms of both the phase velocity and the attenuation. This considerable difference is probably attributable to the hole correction occurring in the CN model. As mentioned above, the simulation at $\phi=6\%$ was performed using only three realizations of the simulation domain, which resulted in the non-smooth RDF shown in figure \ref{fig:distrib}; however, the CN and WT models both gave excellent results at $\phi=6\%$. This indicates that the hole correction has the most significant effects at $\phi \gtrsim 10\%$. At $\phi \geq 24\%$, there is a marked discrepancy between the RDF and the Heaviside step function. A more realistic form of $p({\bf r}_2|{\bf r}_1)$ presented in (\ref{eq:EnsAverage}) might extend the range of validity of the CN model to include higher densities.

\begin{figure}[htbp]
\begin{center}
\begin{tabular}{cc}
Waterman-Truell ($L$ wave) & Waterman-Truell ($T$ wave)\\
\includegraphics[width=6.6cm]{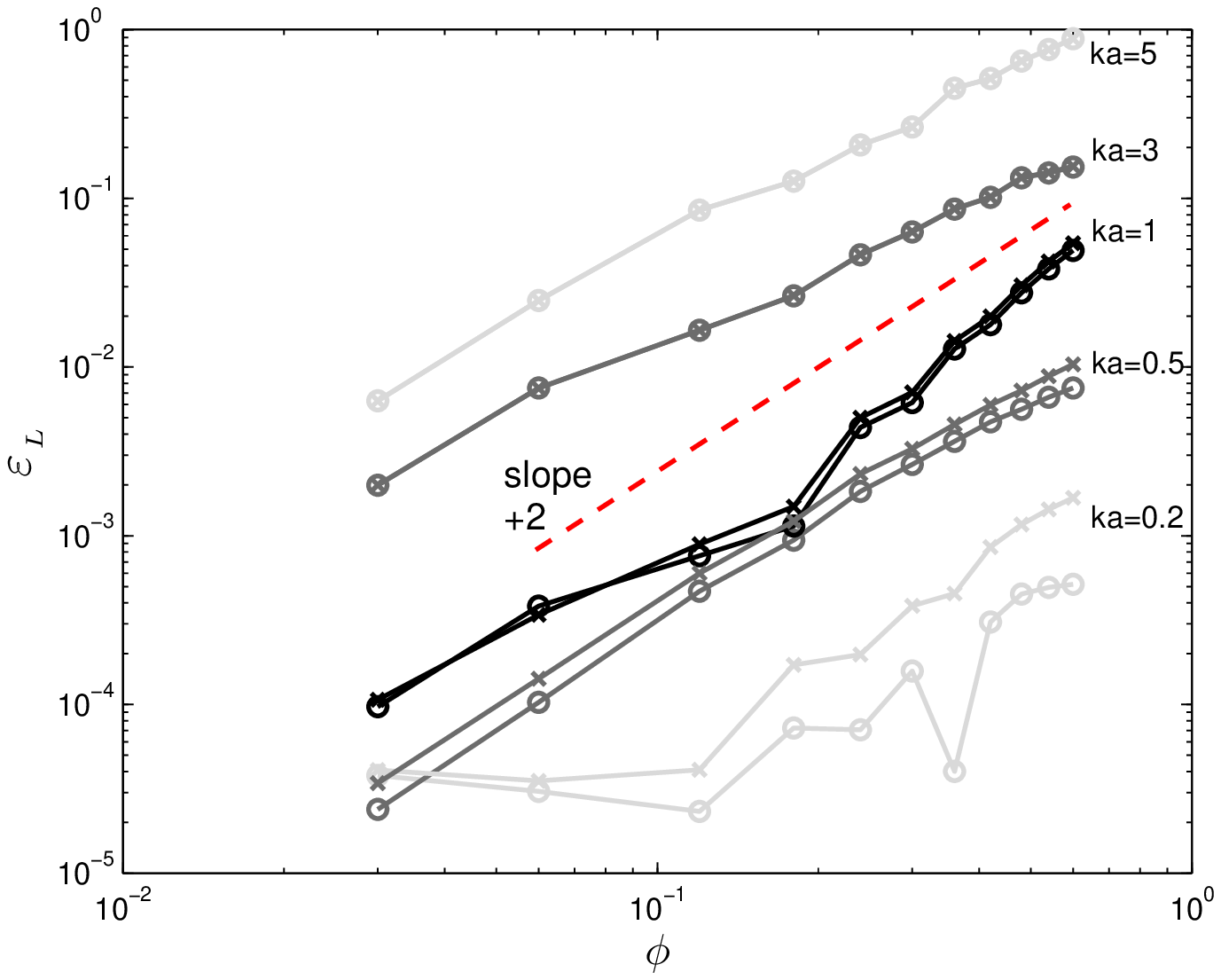}&
\includegraphics[width=6.6cm]{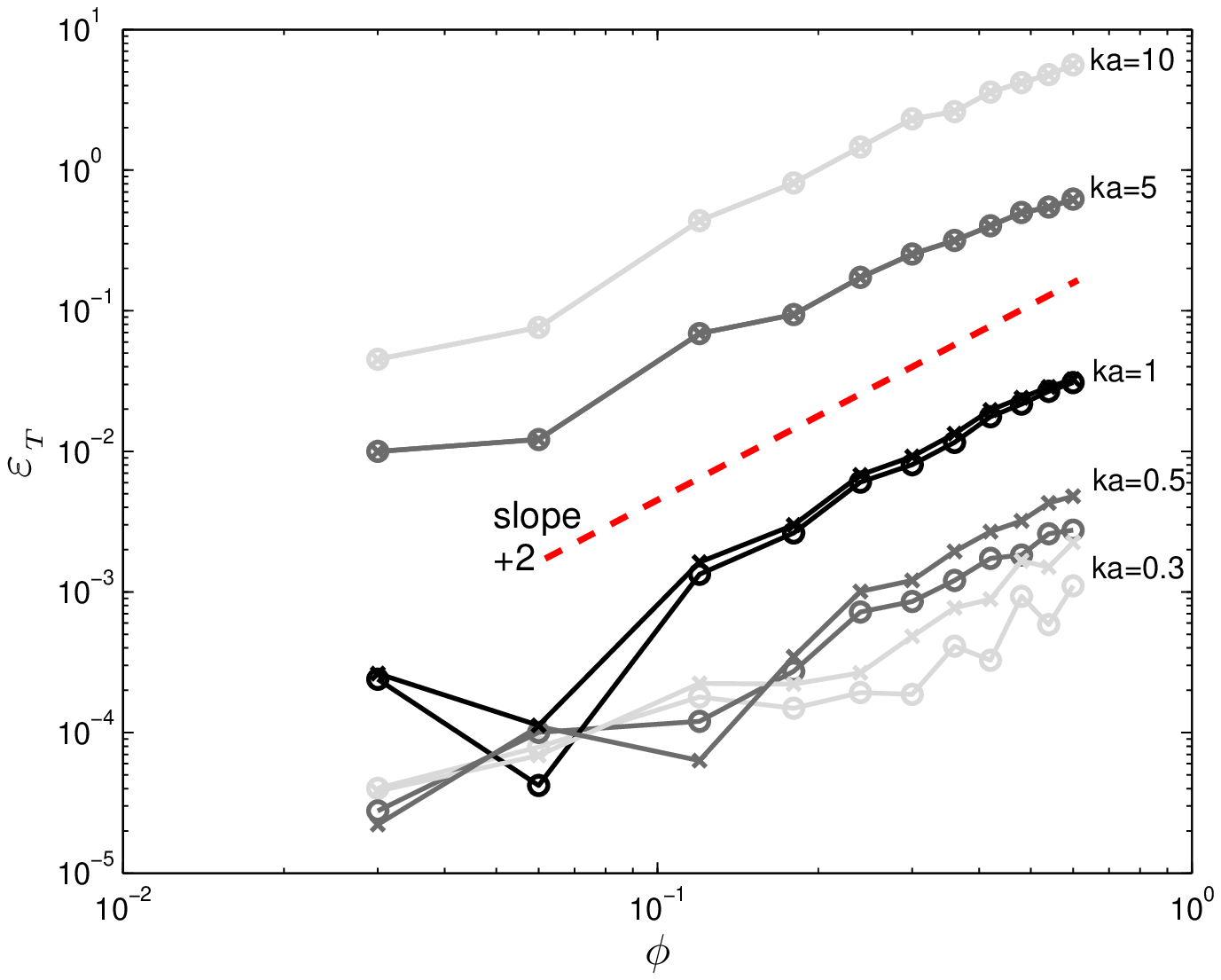}\\
Conoir-Norris ($L$ wave) & Conoir-Norris ($T$ wave)\\
\includegraphics[width=6.6cm]{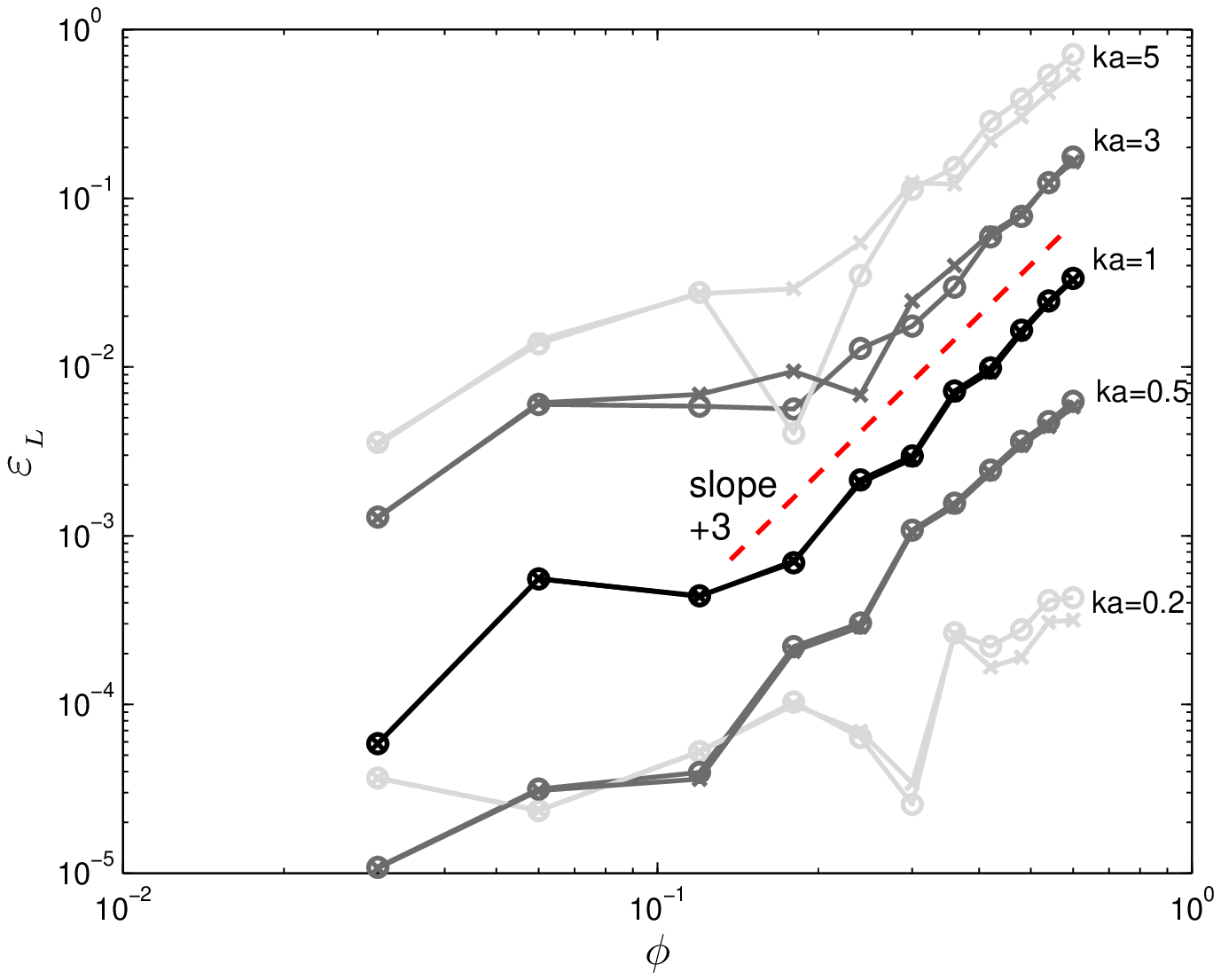}&
\includegraphics[width=6.6cm]{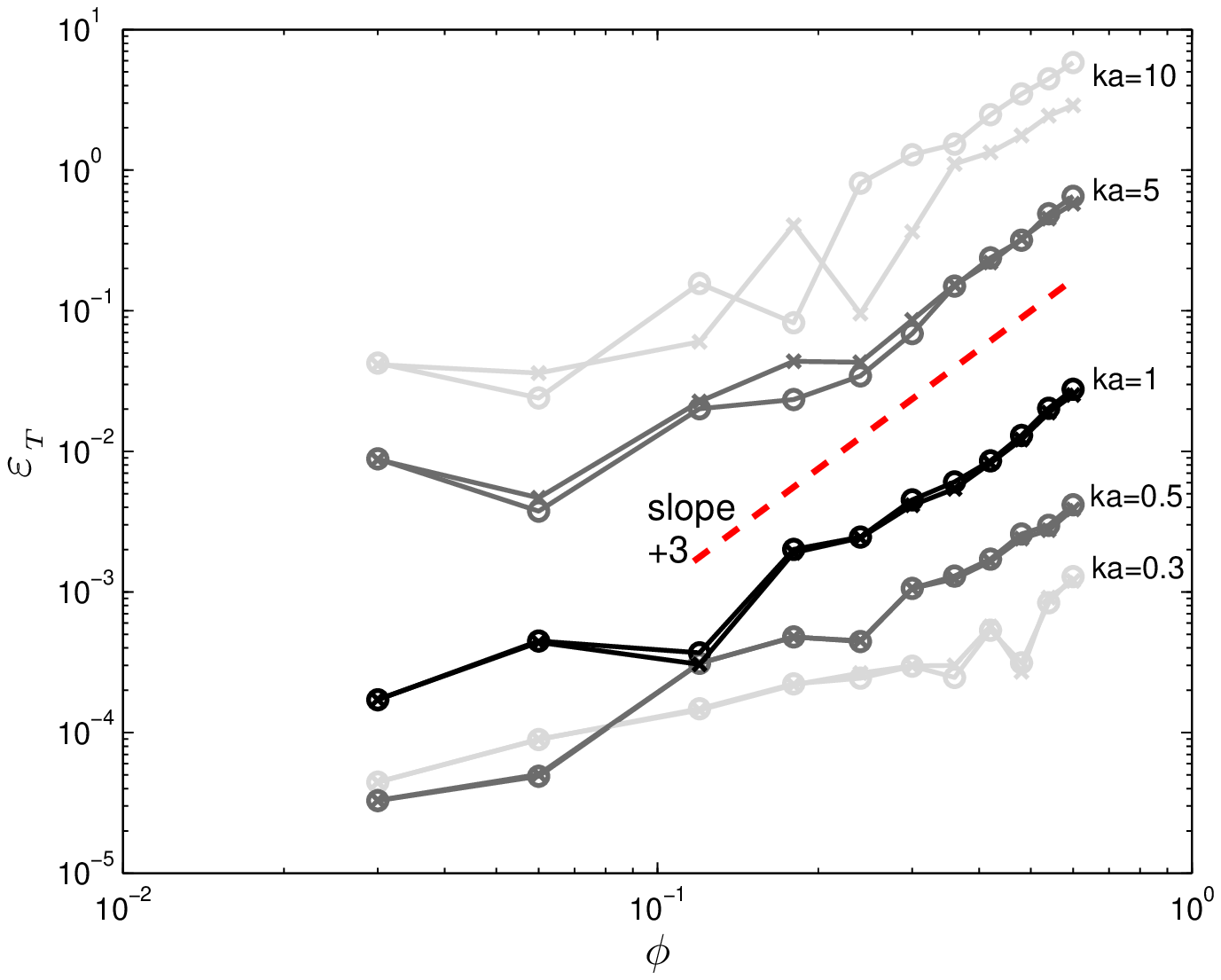}
\end{tabular}
\end{center}
\caption{Error $\varepsilon_\beta$ ($\beta=L,T)$ versus $\phi$ at various dimensionless frequencies (top: WT; bottom: CN; left: longitudinal wave; right: transverse wave). $\times$: implicit method; $\circ$: explicit method.}
\label{fig:distance}
\end{figure}

The WT and CN models are often referred to as second-order models, because of the second-order Taylor expansion (\ref{keLOWn0}) used to estimate $k_{e,\beta}$. However, to our knowledge, this definition has never been justified either theoretically or experimentally. We will therefore examine this question numerically. Let us assume
\begin{equation} 
{\tilde k}_{e,\beta}^2=k_{0,\beta}^2+n_0\,d_{1,\beta}+(n_0)^2\,d_{2,\beta}+\mathcal{O}(n_0^3),
\end{equation}
where ${\tilde k}_{e,\beta}$ is the effective wavenumber obtained in the numerical simulations. The difference between the numerical and the theoretical results are defined by
\begin{equation}
\begin{array}{lll}
\varepsilon_\beta&=&
\displaystyle
\|({\tilde k}_{e,\beta}\,a)^2-({\overline k}_{e,\beta}\,a)^2\|,\\
[8pt]
&=& \displaystyle
a^2\|n_{0}\left(\delta_{1,\beta}-d_{1,\beta}\right)+(n_{0})^2\left(\delta_{2,\beta}-d_{2,\beta}\right)\|+ \mathcal{O}(n_{0}^3),
\end{array}
\end{equation}
where ${\overline k}_{e,\beta}$ is the wavenumber deduced from the WT or CN model. This difference is plotted at various dimensionless frequencies versus the concentration $\phi=n_{0}\,\pi a^2$ in figure \ref{fig:distance}. With the WT model, $\varepsilon_{\beta}$ is governed throughout the whole frequency and concentration range by a slope of 2 decades per decade on a log-log scale: $\varepsilon_{\beta}$  therefore depends mainly on $\delta_{2,\beta}-d_{2,\beta}$. In other words, the WT model predicts $d_{2,\beta}$ inaccurately, even at low densities. In fact, there is no reason why the WT model should be preferred to the ISA model, which is only a first-order model. With the CN model, $\varepsilon_{\beta}$ is independent of the inclusion concentration up to $\phi\approx 18\%$, and has a +3 slope on log-log scale. This confirms the accuracy of the second-order coefficient $d_{2,\beta}$ obtained with the hole correction.

When $\varepsilon_\beta$ was calculated using an implicit formulation to obtain the wavenumber (\ref{equ:SystMod}), $\varepsilon_\beta$ was found to have the same concentration-dependent properties with both the WT and the CN models ($\times$-curves in figure \ref{fig:distance}). The implicit formulation did not improve the quality of the results, and CN is therefore an intrinsically-second-order model.
 

\section{Conclusion}\label{SecConclu}

The effective properties of random elastic media were calculated here using purely numerical methods. Combining sophisticated methods of simulation (the fourth-order ADER scheme and the immersed interface method) and signal processing tools yielded reference solutions for both the real and imaginary parts of the effective wavenumbers. With this approach, the accuracy of the simulations does not depend on the scatterer concentration. Maximum computational efficiency is obtained by performing domain decomposition and parallelizing the algorithms.

This numerical method was applied in the present study to a 2-D model of concrete. In this case, the numerical simulations confirmed that traditional models (such as the Waterman-Truell model) are valid roughly up to inclusion densities of 10\%, whereas the recent Conoir-Norris (extension of Linton-Martin to elastodynamics) is valid up to densities of 25 \%. In particular, the present simulations confirmed the validity of the second-order term in the Conoir-Norris model, as previously done in a theoretical study \cite{Conoir10}. We hope that a similar approach can be used by researchers to test the validity of their favorite multiple-scattering model \cite{Kanaun03}.

The numerical method presented here can be used to handle more complex configurations, such as a granulometry or composite containing scatterers of various shapes. Other constitutive laws could also be introduced, such as viscoelastic laws accounting for dissipative effects \cite{Lombard11}. Lastly, the possibility of extending the present approach to 3-D configurations is a great computational challenge. Preliminary tests have already been conducted on the numerical methods with fluid scatterers included in a fluid matrix.


\section*{Acknowledgements}

The authors wish to thank Dr Emilie Franceschini for her helpful comments, and Jessica Blanc for her careful reading of the manuscript.

\appendix

\section{Effective wavenumbers}\label{AppEff}

As established in \cite{Varadan80,Conoir10}, the scattered field $\psi_{s}$ recorded at ${\bf r}$ and due to an inclusion centered at ${\bf r_{1}}$ can be expressed through a linear operator ${\cal T}$ applied to the exciting field $\psi_{e}$: $\psi_{s}({\bf r};{\bf r_{1}}) = {\cal T}({\bf r_{1}}) \ \psi_{e}({\bf r};{\bf r_{1}})$. If the positions of the $N$ inclusions are known, closed-form solution of the problem can be obtained. If, on the contrary, the inclusions are randomly distributed, it is generally attempted to determine the effective field $\left< \psi_{e}\right>$ corresponding to the ensemble average of the positions of all the inclusions. The effective field at one representative inclusion (say the first one) is expressed in terms of the scattering induced by another representative inclusion (say the second). Based on the quasicristalline approximation (QCA) \cite{Lax52}, the latter scattered field is assumed to be excited by the same effective field as the first inclusion:
\begin{equation}
\left< \psi_{e}({\bf r};{\bf r}_{1})\right>=\psi_{i}({\bf r})+
(N-1)\int {\cal T}({\bf r_{2}})\left< \psi_{e}({\bf r};{\bf r}_{2})\right>
p({\bf r_{2}}|{\bf r_{1}}) \,d{\bf r_{2}},
\label{eq:EnsAverage}
\end{equation}
where $\psi_{i}$ is the incident field. The probability density $p({\bf r_{2}}|{\bf r_{1}})$ in (\ref{eq:EnsAverage}), which is a pair-correlation function, expresses the probability of finding an inclusion at ${\bf r_{2}}$, given that an inclusion is placed at ${\bf r_{1}}$.

Various models have been developed for determining the effective wavenumbers analytically, such as the Waterman-Truell (WT) \cite{WatermanTruell61} and Lloyd-Berry (LB) \cite{Lloyd67} models, to cite but a few. As pointed out by Linton-Martin \cite{Linton05}, these models differ mainly in the hole correction, i.e. the assumption made about $p$ in the integration of (\ref{eq:EnsAverage}). The LB model is based on the following assumption:
\begin{equation}
\frac{N-1}{n_{0}}\,p({\bf r_{2}}|{\bf r_{1}}) =
\left\{
\begin{array}{ll}
0   &\quad \mbox{if }|{\bf r_{2}}-{\bf r_{1}}|< b,\\
[8pt]
1	&\quad \mbox{if }|{\bf r_{2}}-{\bf r_{1}}|> b,
\end{array}
\right.
\quad\quad b=2\,a+\xi,
\label{eq:formeQCA}
\end{equation}
where $a$ and $\xi$ are defined in section \ref{SecRandomAlgo}. To obtain the effective wavenumbers in $L$ and $T$ waves, the effective field can be decomposed into a modal sum, where the modal amplitude depends on the effective wavenumber \cite{Conoir10}. Introducing this form into (\ref{eq:EnsAverage}) and using (\ref{eq:formeQCA}), the problem reduces to searching for the non-trivial solution of the infinite linear system \cite[equation (31)]{Conoir10}:
\begin{equation}
\det\left({\bf I} - 2\,n_{0}\,{\bf M}\;{\bf T}\right)=0,
\label{equ:SystMod}
\end{equation}
where ${\bf I}$ is the identity matrix, $\bf T$ is a matrix defined in \cite[equation (37)]{Conoir10}, and 
\begin{equation}
\begin{array}{l}
{\bf M}=
\left(
\begin{array}{ll}
\displaystyle {\bf M}_L	& {\bf 0} \\
{\bf 0}		&\displaystyle {\bf M}_T
\end{array}
\right),\\
\\
\displaystyle
{\bf M}_{\beta}[m,n]=\,\frac{\pi}{k_{e}^2-k_{\beta}^2}\left(k_{e}\,b\,J'_{m-n}(k_{e} b)\,H^{(1)}_{m-n}(k_{\beta} b)-k_{\beta}\,b\,J_{m-n}(k_{e}b)\,H^{(1)}_{m-n}\,'(k_\beta b) \right).
\end{array}
\end{equation}
In practice, the modal sum in (\ref{equ:SystMod}) is truncated. The effective wavenumbers $k_{e}$ satisfying (\ref{equ:SystMod}) are associated with $k_{e,L}$ and $k_{e,T}$. However, searching $k_{e,L}$ and $k_{e,T}$ is an intricate and time-consuming process. Another explicit but approximate form can be obtained using Taylor expansions:
\begin{equation}
k_{e,\beta}^2=k_{\beta}^2+n_{0}\,\delta_{1,\beta}+(n_{0})^2\,\delta_{2,\,\beta},
\label{keLOWn0}
\end{equation}
where $\delta_{1,\beta}$ and $\delta_{2,\beta}$ are defined in \cite[equations (62a) and (62b)]{Conoir10} in the Conoir-Norris (CN) model. Let us now consider the WT model, which is based on
\begin{equation}
\frac{N-1}{n_{0}}\,p({\bf r_{2}}|{\bf r_{1}}) =
\left\{
\begin{array}{ll}
0		&\quad \mbox{if } \left(\bf{r_{2}}-\bf{r_{1}}\right)\cdot\bf{e}_{y}< \eta,\\
[8pt]
1	&\quad \mbox{if }\left(\bf{r_{2}}-\bf{r_{1}}\right)\cdot\bf{e}_{y}> \eta,\\
\end{array}
\right.
\quad\quad \text{with}\quad \eta\to 0.
\label{eq:formeWT}
\end{equation}
where $\left(\bf{r_{2}}-\bf{r_{1}}\right)\cdot\bf{e}_{y}$ is the distance between two inclusions in the direction of propagation ($\bf{e_{y}}$ here) of the incident wave. Based on this hypothesis, the system will have the same form as (\ref{equ:SystMod}), but $\bf M$ will be different, and this parameter is deduced from \cite[equations (12-13)]{Luppe10}:
\begin{equation}
{\bf M}_{\beta}[m,n] = \frac{1}{i\,k_\beta}\left(\frac{1}{k_{e}-k_\beta}-\frac{(-1)^{m-n}}{k_{e}+k_\beta}\right).
\end{equation}
The approximate form (\ref{keLOWn0}) can still be used with the WT model. The same expression for $\delta_{1,\beta}$ in (\ref{keLOWn0}) can be used as with the LB model, but the definition for $\delta_{2,\beta}$ in this case is that given in \cite[equation (4)]{Linton05}. Note that ISA matches with (\ref{keLOWn0}) to the first order: $k_{e,\beta}^2=k_{\beta}^2+n_{0}\,\delta_{1,\beta}$. 


\end{document}